\documentclass[preprint]{aastex62}

\usepackage{graphicx, epstopdf, natbib, booktabs, bigstrut, enumerate}
\graphicspath{{./figures/}}
\usepackage{threeparttable, enumerate, amsmath}
\usepackage{rotating, multirow, gensymb}

\received{\today}
\submitjournal{ApJ}

\shortauthors{Dogruel et al.}

\begin{document}

\title{X-RAY MONITORING OF GRAVITATIONALLY LENSED RADIO-LOUD QUASARS WITH \textit{CHANDRA}}

\correspondingauthor{Mustafa Burak Dogruel}
\email{burak@ou.edu}

\author{Mustafa Burak Dogruel}
\affil{Homer L. Dodge Department of Physics and Astronomy, The University of Oklahoma, 440 W. Brooks St. Norman, OK 73019, USA}

\author{Xinyu Dai}
\affil{Homer L. Dodge Department of Physics and Astronomy, The University of Oklahoma, 440 W. Brooks St. Norman, OK 73019, USA}

\author{Eduardo Guerras}
\affil{Homer L. Dodge Department of Physics and Astronomy, The University of Oklahoma, 440 W. Brooks St. Norman, OK 73019, USA}

\author{Matthew Cornachione}
\affil{Department of Physics, United States Naval Academy, 572C Holloway Road, Annapolis, MD 21402, USA}

\author{Christopher W. Morgan}
\affil{Department of Physics, United States Naval Academy, 572C Holloway Road, Annapolis, MD 21402, USA}



\begin{abstract}

	In this work, we calculated the sizes of unresolved X-ray emission regions in three gravitationally lensed radio-loud quasars, B\,1422+231, MG\,J0414+0534 and Q\,0957+561, using a combination of imaging and spectral analysis on the X-ray data taken from the \textit{Chandra X-Ray Observatory}. We tentatively detected FeK$\alpha$ emission lines in MG\,J0414+0534 and Q\,0957+561 with over 95\% significance, whereas, we did not significantly detect FeK$\alpha$ emission in B\,1422+231. We constructed differential microlensing light curves from absorption corrected count rates. We subsequently performed a microlensing analysis on the X-ray microlensing light curves to measure the X-ray source sizes in soft (0.83--3.6 keV), hard (3.6--21.8 keV), and full (0.83--21.8 keV) bands, based on either Bayesian or maximum likelihood probabilities. For B\,1422+231, sizes from the two methods are consistent with each other, e.g. $R_X^{hard}/R_G = 6.17\pm5.48 \text{ (Bayesian), } 11.81\pm3.75 \text{ (maximum likelihood)}$, where $R_G=GM_{BH}/c^2)$. However, for MG\,J0414+0534 and Q\,0957+561, the two methods yield completely different results suggesting that more frequently sampled data with better signal-to-noise ratio are needed to measure the source size for these two objects. Comparing the acquired size values with the radio-quiet sample in the literature we found that our results are consistent with X-ray source size scaling approximately as $R_X \propto M_{BH}$ with the mass of the central supermassive black hole. Our results also indicate that radio-loud quasars tend to have larger unresolved X-ray emission sizes compared to the radio-quiet ones.
	
\end{abstract}




\section{Introduction} \label{intro}
Unification schemes of active galactic nuclei (AGNs) have indicated that AGNs are separated into two physically distinct classes, radio-loud and radio-quiet \citep{wilson1995, urry1995}, where the radio-loudness is caused by the presence of relativistic jets. Depending on redshift and luminosity, radio-loud AGNs constitute roughly $\sim 4-25\%$ of AGN population \citep{keller1989,jiang2007}. The relativistic radio jets of these radio-loud AGNs have also been observed in X-rays, which was a surprising discovery of \textit{Chandra} based on early jet models, e.g., PKS 0637--752 \citep{schwartz2000, chartas2000}. The fact that many of these jets can also be easily detected in X-rays means that the X-ray emission from radio-loud quasars emanates not only close to the accretion disc, as the radio-quiet counterparts, but also from the jets. The resolved X-ray emission from radio-loud quasars is associated with kpc-scale jets \citep[e.g.][]{chartas2000,marshall2018}, whereas the unresolved X-ray emission from radio-loud quasars is still not clear. This elusiveness creates a major challenge in interpreting the properties of quasar continuum in X-rays for radio-loud quasars. The unresolved component of X-ray emission is thought to be a combination of corona emission, resembling the case of radio-quiet AGNs, and the contribution from the unresolved jet. Measuring the spatial extent of the unresolved X-ray emission in radio-loud quasars and comparing that with the measurements of radio-quiet quasars will provide an additional constraint on separating the jet and corona contributions. For this purpose, quasar microlensing phenomenon provides one of the strongest methods.

AGNs have a critical role in cosmic evolution. For instance, observations of $z>6$ quasars constrain the formation of the first super massive black holes in the early universe. Furthermore, the existence of tight correlations between the super massive black hole mass and host galaxy properties, luminosity, mass and velocity dispersion $(\sigma)$ of the stellar bulge/spheroid, \citep[e.g.][]{kormendy1995, fer2000, mccon2013} shows that these black holes regulate galaxy evolution and vice versa. Powered by the central super massive black hole, AGN feedback is an indispensable component in modeling galaxy evolution \citep{somer2008}. Despite these crucial aspects, the structure of AGNs is not yet fully understood. For radio-quiet quasars, the thin disc model does not predict X-ray emission for massive AGNs, and the emission is expected from a corona \citep{blaes2007}. One of the biggest problems in testing accretion disc models is that the central engine of AGNs cannot be resolved even with space telescopes \citep{mosq2013}. For instance, according to some rough estimates, the angular size of the central engine is of the order of nano-arcseconds \citep{dai2010}. 

Quasar microlensing is induced by the joint lensing of an ensemble of stellar mass objects in a foreground galaxy between the observer and the quasar. The technique has been proven to be an efficient way of probing the innermost regions of AGNs \citep[e.g.][]{dai2010, mosq2013, blackburne2014}. Since the quasar, the lens galaxy and the stars within it, and the observer have relative motion transverse to the line of sight \citep{wamb2006}, the angular location of the quasar relative to the lens galaxy changes with time. Thus, the magnification of each image of the quasar varies due to microlensing, which leads to uncorrelated flux variations between the lensed images. The microlensing magnifications also depend on the relative sizes of the emission region (here the accretion disc of the quasar) and also on the Einstein radius of the star, which can be approximated for a cosmological lens as
\begin{equation}
	R_E = \sqrt{\frac{4GM}{c^2}\frac{D_{os} D_{ls}}{D_{ol}}} \sim 9\times 10^{16} cm \sqrt{\frac{M}{M_\odot}} \sqrt{\frac{D_{os}}{c/H_0} \frac{D_{ls}}{D_{ol}}}
\label{einsteinrad}
\end{equation}
where $M$ is the mass of the deflector, $D_{os}, D_{ls}, D_{ol}$ are the angular diameter distances between the observer, lens and the source respectively, and $c/H_0$ is the Hubble radius. This dependence implies that the smaller the source size, the greater the microlensing amplitude, which means that the amplitude of the microlensing variations can be used to measure the source size. 

The largest microlensing amplitudes are observed in X-rays \citep{chartas2002, dai2003, mosq2013}. The UV photons emitted from the inner regions of accretion disc undergo inverse Compton scattering by the relativistic electrons in the corona to produce X-ray continuum which can be characterised by a power law. Since electron scattering is isotropic, some of these photons are scattered back to the disc, forming the reflection component which can also include emission features such as the FeK$\alpha$ fluorescent line (the strongest of those emission lines) at 6.4 keV in the rest frame \citep{georgefabian1991, fabian1995, gou2011}. Studying the gravitational microlensing of X-rays from quasars provides us with an opportunity to estimate the size of the X-ray emitting region of the accretion disc. Even though gravitationally lensed quasars are quite few in numbers, they provide a powerful and effective tool to probe the inner structure of quasars which cannot be resolved spatially by telescopes. Another benefit of microlensing analysis is that it can be used to measure the innermost stable circular orbit of the central supermassive black holes which makes it possible to constrain the spin of the black holes \citep{dai2019}. Furthermore, microlensing analysis can constrain the discrete lens population including extragalactic planets \citep{dai2018}.

In this study, we present the X-ray spectra and light curves for three gravitationally lensed radio--loud quasars MG\,J0414+0534, Q\,0957+561, and B\,1422+231. We extract the full (0.83 -- 21.8 keV rest frame), soft (0.83 -- 3.6 keV), and hard (3.6 -- 21.8 keV) X-ray band light curves and compare them with image flux ratio predictions without microlensing to measure the microlensing signals. We model the microlensing variability and then generate a probability density function (PDF) to constrain the size of the unresolved X-ray emitting region of the aforementioned three radio-loud quasars. Finally, we discuss the results in Section \ref{conc}. Throughout the paper, we assume a flat $\Lambda$CDM cosmology with $H_0=70\text{ km s}^{-1}\text{ Mpc}^{-1}, \Omega_m=0.3$ and $\Omega_\Lambda=0.7$.

\section{Observations and Data Analysis} \label{obs}

\begin{figure}
	\centering
	\includegraphics[scale=0.36]{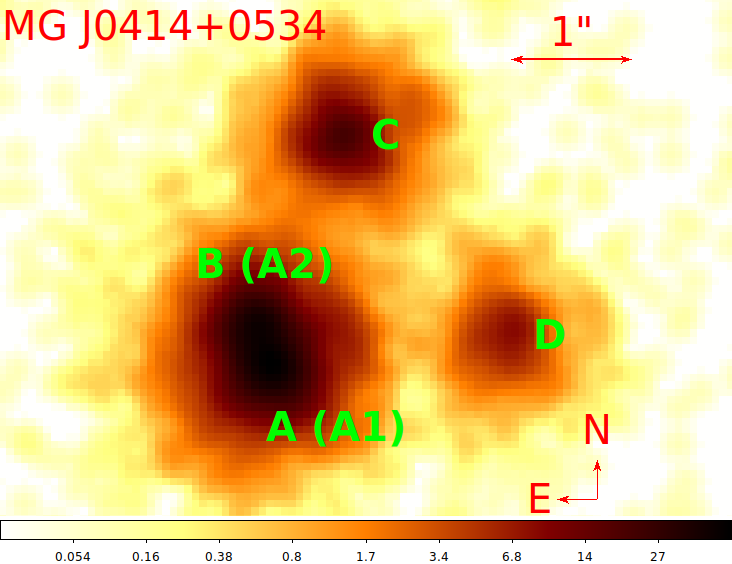}
	\includegraphics[scale=0.36]{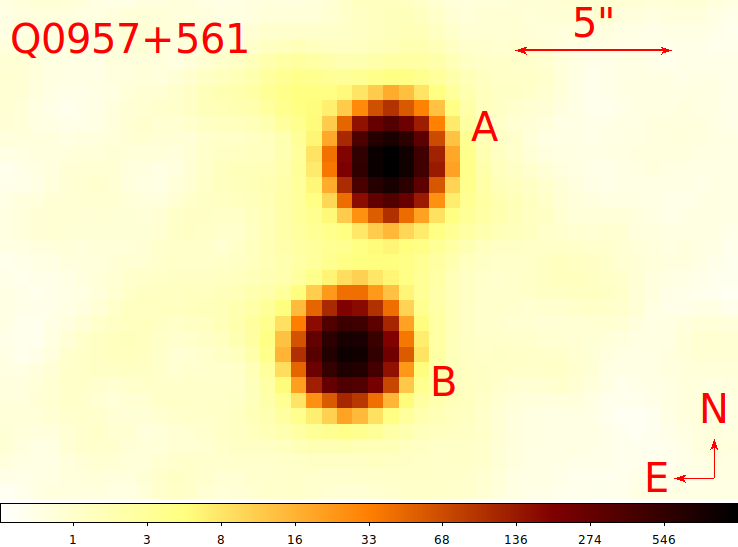}
	\includegraphics[scale=0.36]{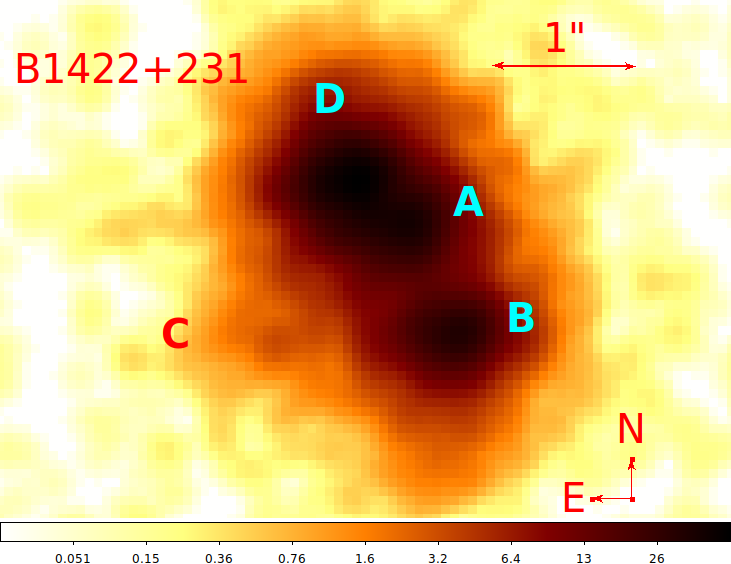}
	\caption{Stacked \textit{Chandra} images of MG\,J0414+0534, Q\,0957+561 and B\,1422+231.}
	\label{stack_obs}
\end{figure}

Observations were performed with the Advanced CCD Imaging Spectrometer on the \textit{Chandra X-Ray Observatory} which has an on-axis point spread function (PSF) of $0\farcs5$. We selected three radio-loud quasars that have multi epoch observations in the \textit{Chandra} Data Archive\footnote{\url{http://cda.harvard.edu/chaser/}} and yielded three lenses with their properties listed in Table \ref{obj_tab}. Stacked \textit{Chandra} images of the three targets are shown in Figure \ref{stack_obs}. All data were reprocessed using CIAO 4.7 software\footnote{\url{http://cxc.harvard.edu/ciao/}} tools. 

\begin{table}[htbp]
\resizebox{\textwidth}{!}{
\begin{threeparttable}
  \centering
  \caption{Lens Data For Selected Radio-Loud Quasars}
    \begin{tabular}{lcccccccc}
    \toprule
    \toprule
    \multicolumn{1}{c}{Object} & $z_s$ & $z_l$ & $R_E$  & $t_E$  & $t_{10R_G}$  & $\Delta t_{obs}$  & $M_{BH}$  & $R_G$ \\
          &       &       & (light days) & (years) & (years) & (years) & ($\times 10^9 M_\odot$) & (light days) \\
    \midrule
    MG\,J0414+0534 & 2.64  & 0.96  & 8.054 & 19.39 & 3.08  & 11.75 & 1.82 (C IV) & 0.104 \\
    Q\,0957+561 & 1.41  & 0.36  & 12.788 & 12.39 & 1.11  & 10.19 & 2.01 (C IV) & 0.114 \\
    B\,1422+231 & 3.62  & 0.34  & 12.305 & 23.94 & 4.29  & 11.48 & 4.79 (C IV) & 0.273 \\
    \bottomrule
    \end{tabular}%
  \label{obj_tab}
	\begin{tablenotes}[para, flushleft]
		\small
		\item Source and lens redshifts ($z_s$ and $z_l$) are taken from CASTLES.\\
		\item Einstein radius crossing time ($t_E$) and the mass of the supermassive black hole ($M_{BH}$) are taken from \protect\cite{mosqkoch2011}. $t_{10R_G}$ is $10\,R_G$ crossing time. \\
		\item $R_E$ is calculated assuming a mean stellar mass of $\langle M_\ast \rangle =0.3\,M_\odot$ in lens galaxies.\\
		\item Time span of the observations are given under $\Delta t_{obs}$.\\
		\item Gravitational radius $R_G=GM_{BH}/c^2$, which is half of the Schwarzschild radius $R_S$, is given in the last column.
	\end{tablenotes}
\end{threeparttable}}
\end{table}

\subsection{Imaging Analysis}

We later separated the events into soft and hard bands where the energy boundary was selected to be 3.6 keV in the observed frame to acquire comparable count rates (as given in Tables \ref{corfluxq}--\ref{corfluxb}) between the two energy bands. For all three systems, we subtracted the background emission from image count rates using concentric circular regions with inner and outer radii of $\sim10''$ and $\sim20''$ respectively. Apart from Q\,0957+561, the first gravitationally lensed quasar detected \citep{walsh1979} with well separated images, the angular separation of lensed components of B\,1422+231 and MG\,J0414+0534 can be as small as $0\farcs4$ and $0\farcs5$, respectively. Therefore, it is evidently not suitable to perform aperture photometry since it will be contaminated by the flux of nearby sources in the image. Consequently, to accurately measure the image count rates, we used PSF fitting method with the relative positions of the lensed components which were taken from the CASTLES\footnote{\url{https://www.cfa.harvard.edu/castles/}} database. After the acquisition of background subtracted count rates, they were further corrected for both Galactic absorption and absorption by the lens galaxy measured from the spectral analysis. 

\subsection{Spectral Analysis}

\begin{figure}
	\center
	\includegraphics[scale=0.45]{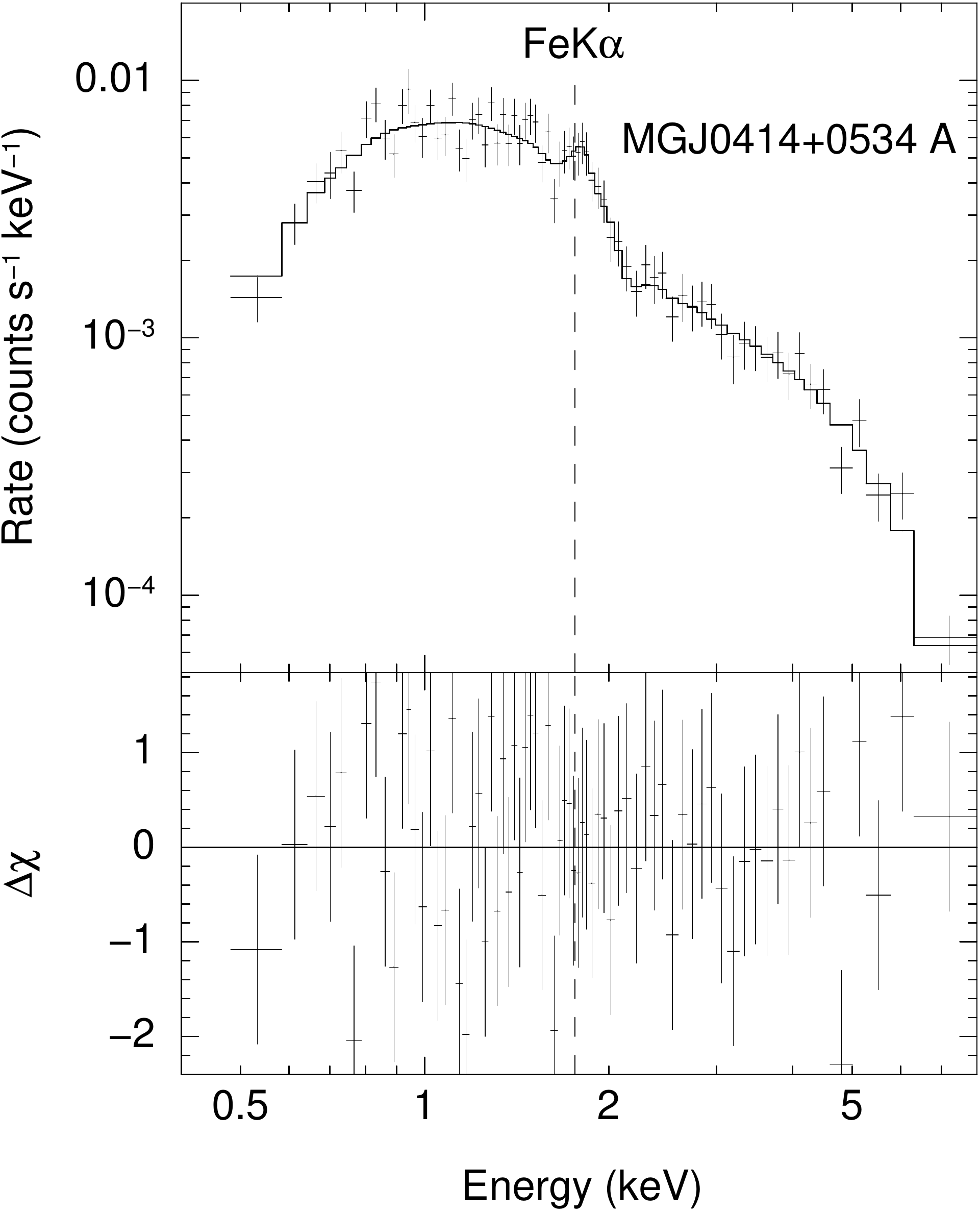} 
	\includegraphics[scale=0.45]{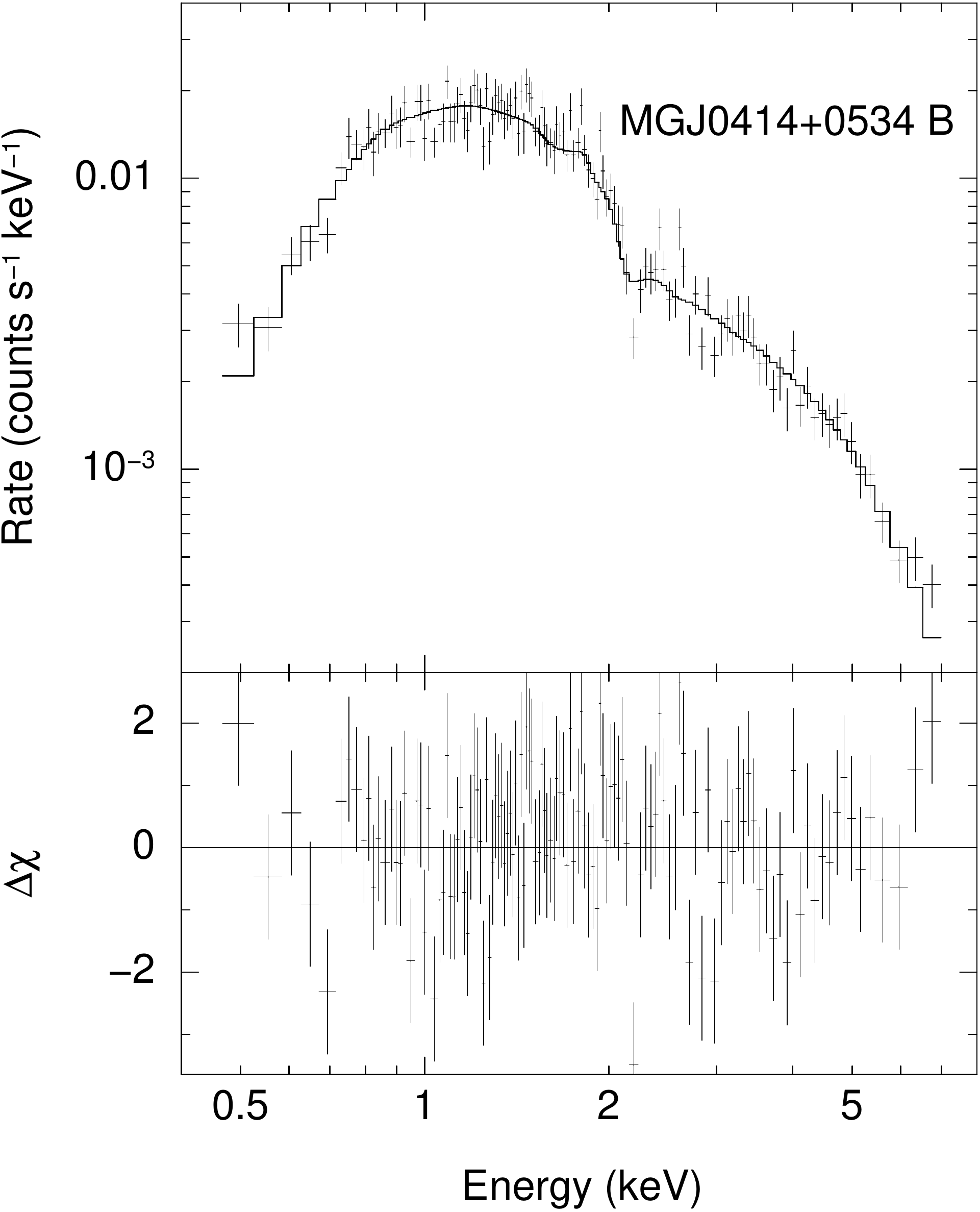}
	\includegraphics[scale=0.45]{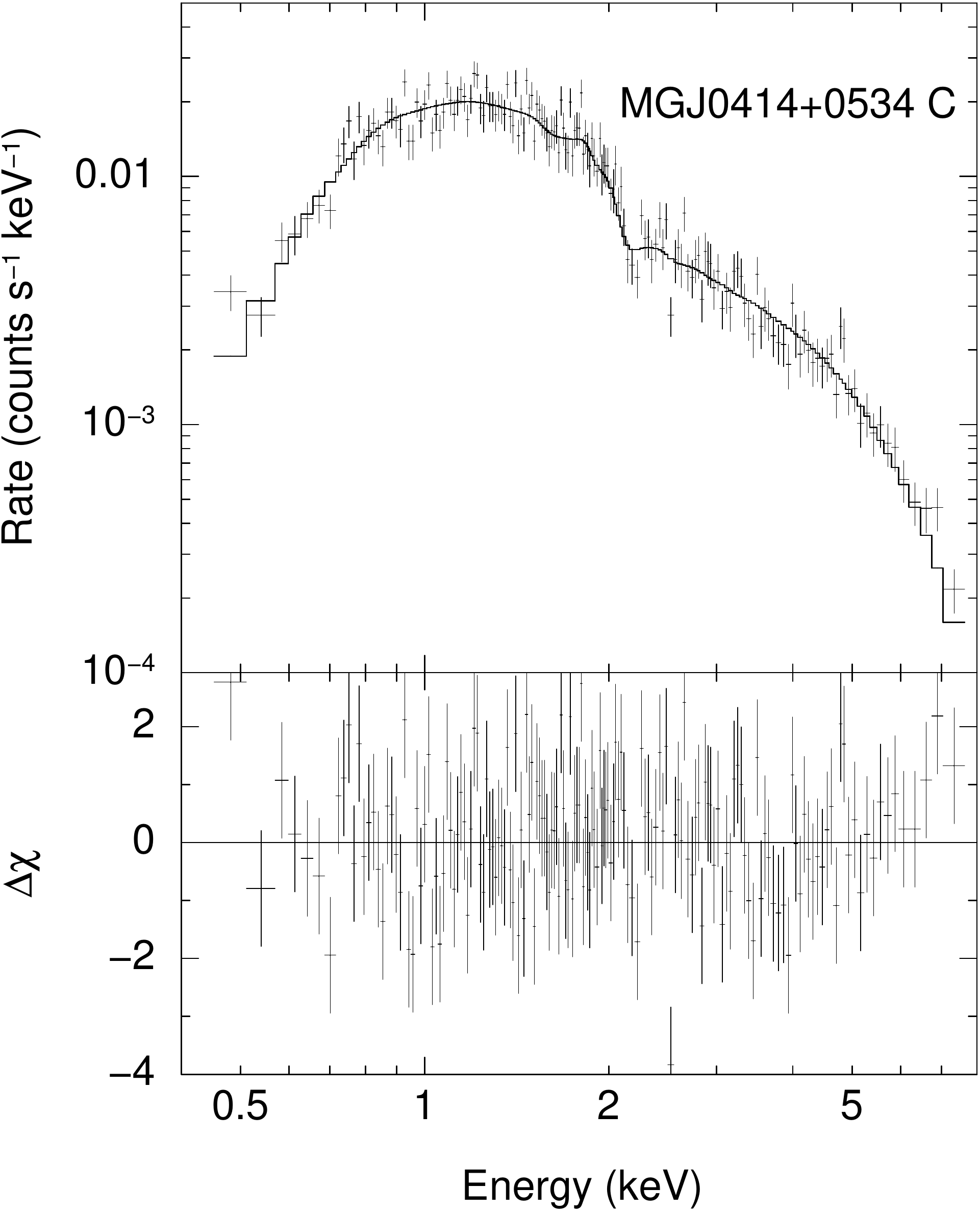} 
	\includegraphics[scale=0.45]{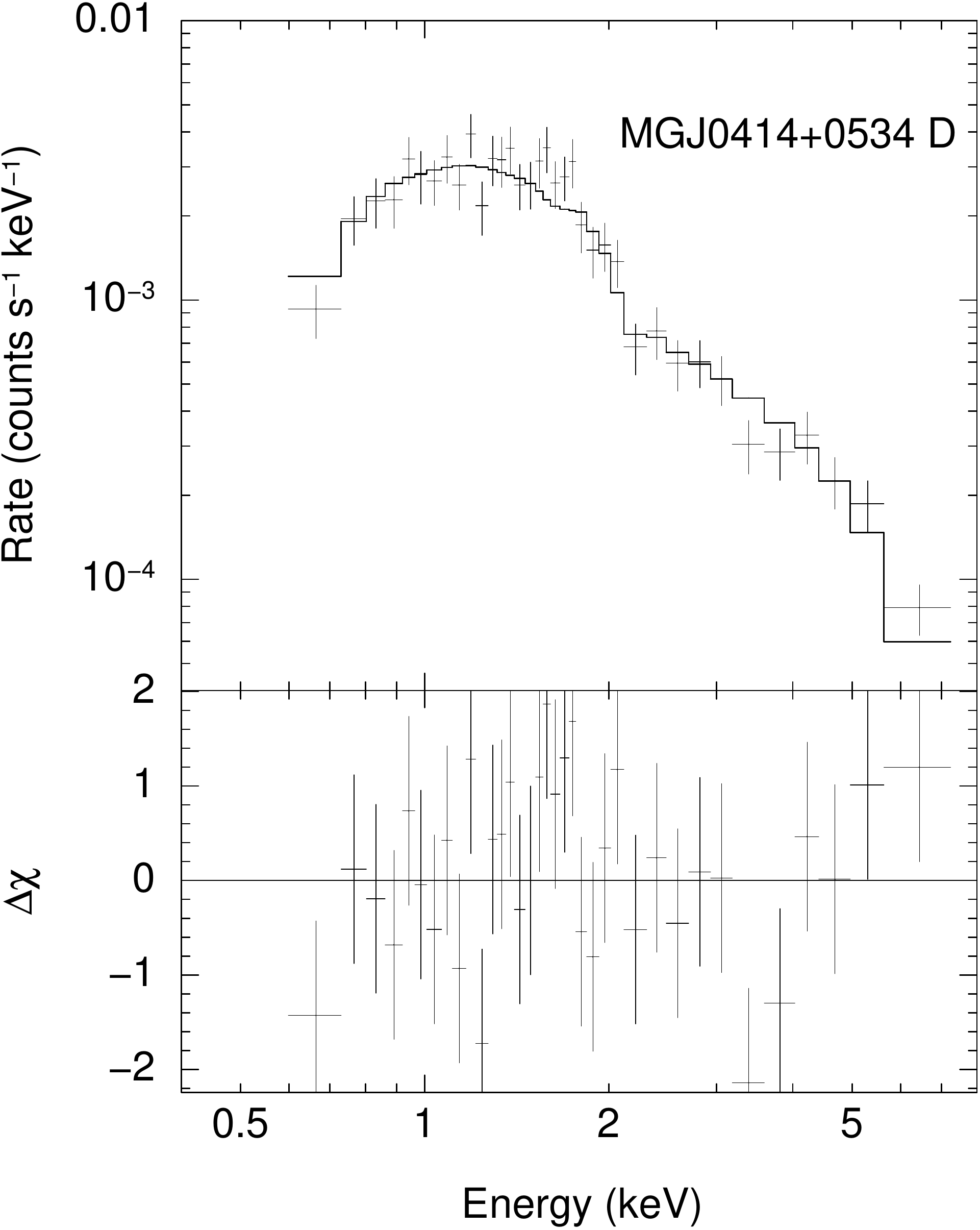}
	\caption{Stacked spectra of MG\,J0414+0534 and spectral fits. The sub-panels show the statistical residuals in units of 1$\sigma$ standard deviations.}
	\label{specfit0414}
\end{figure}

\begin{figure}
	\center
	\includegraphics[scale=0.45]{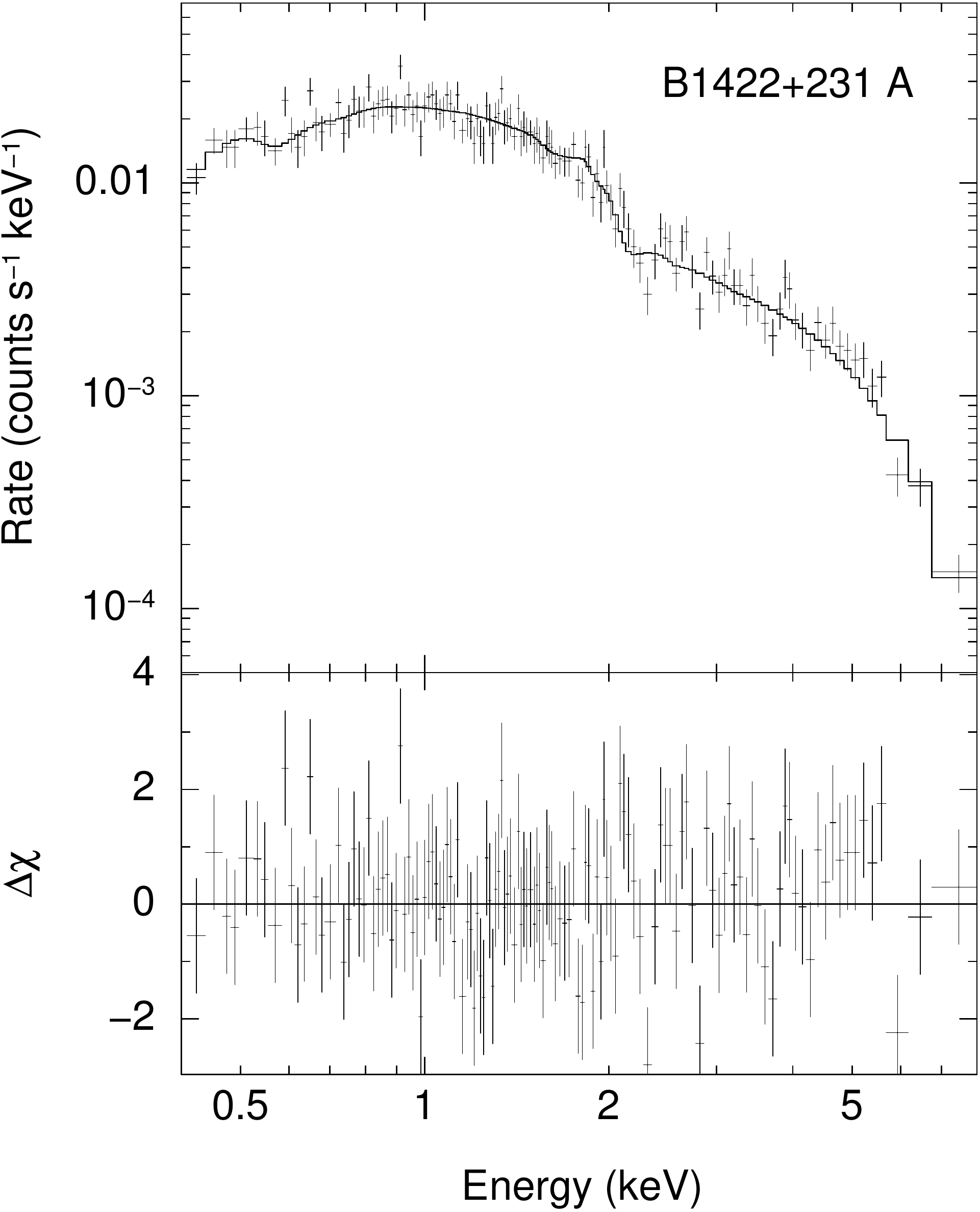} 
	\includegraphics[scale=0.45]{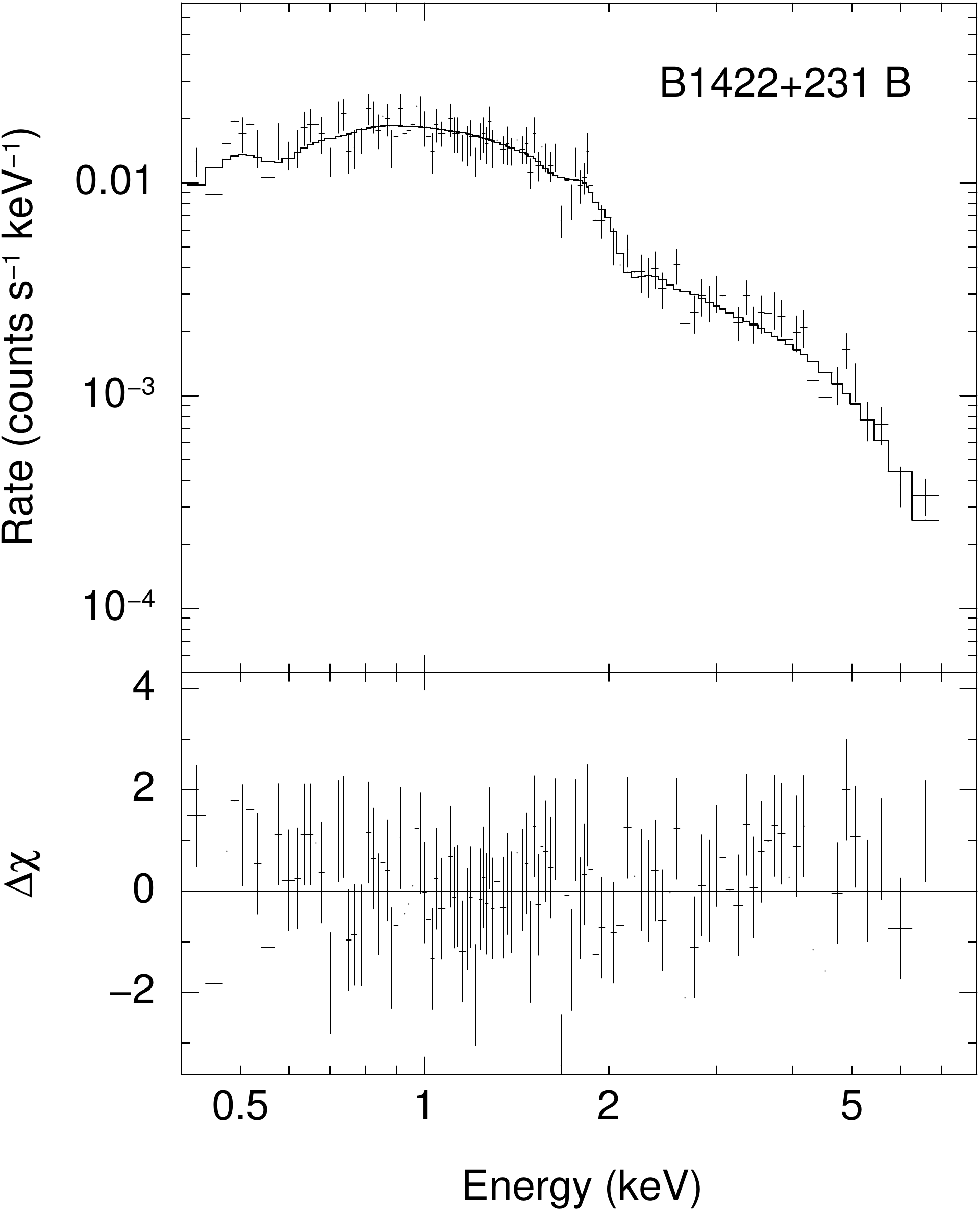}
	\includegraphics[scale=0.45]{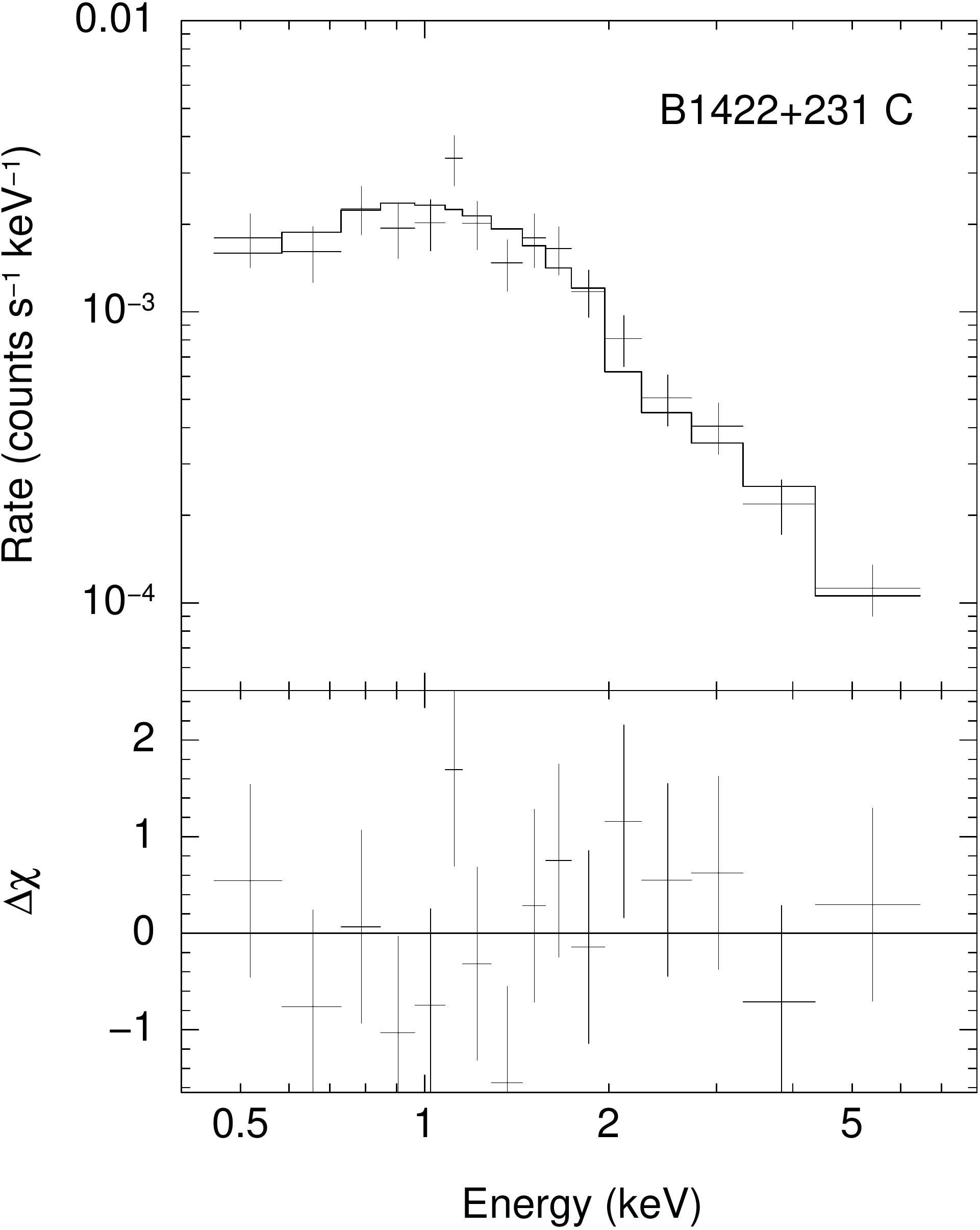} 
	\includegraphics[scale=0.45]{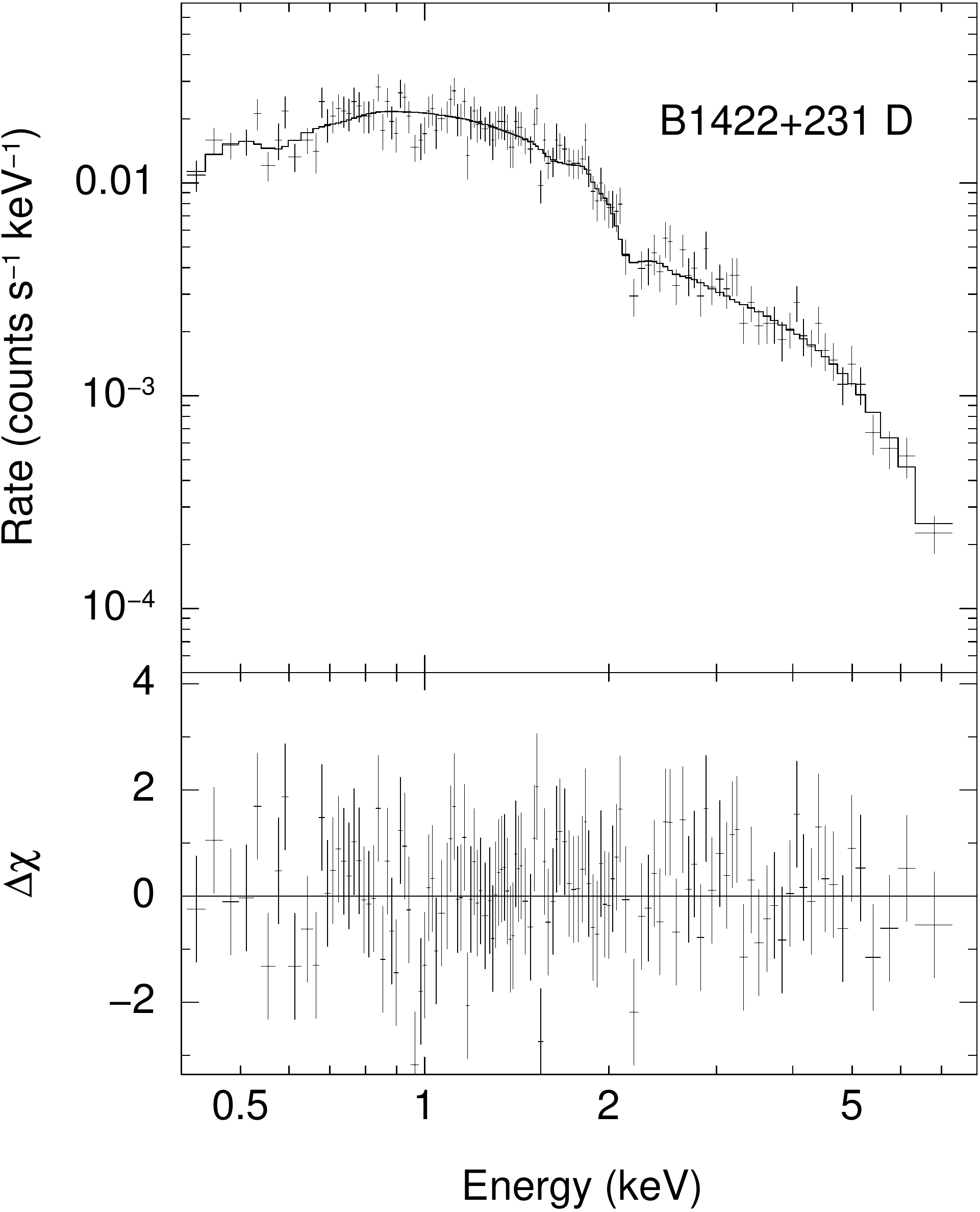}
	\caption{Stacked spectra of B\,1422+231 and spectral fits. The sub-panels show the statistical residuals.}
	\label{specfit1422}
\end{figure}

\begin{figure}
	\center
	\includegraphics[scale=0.44]{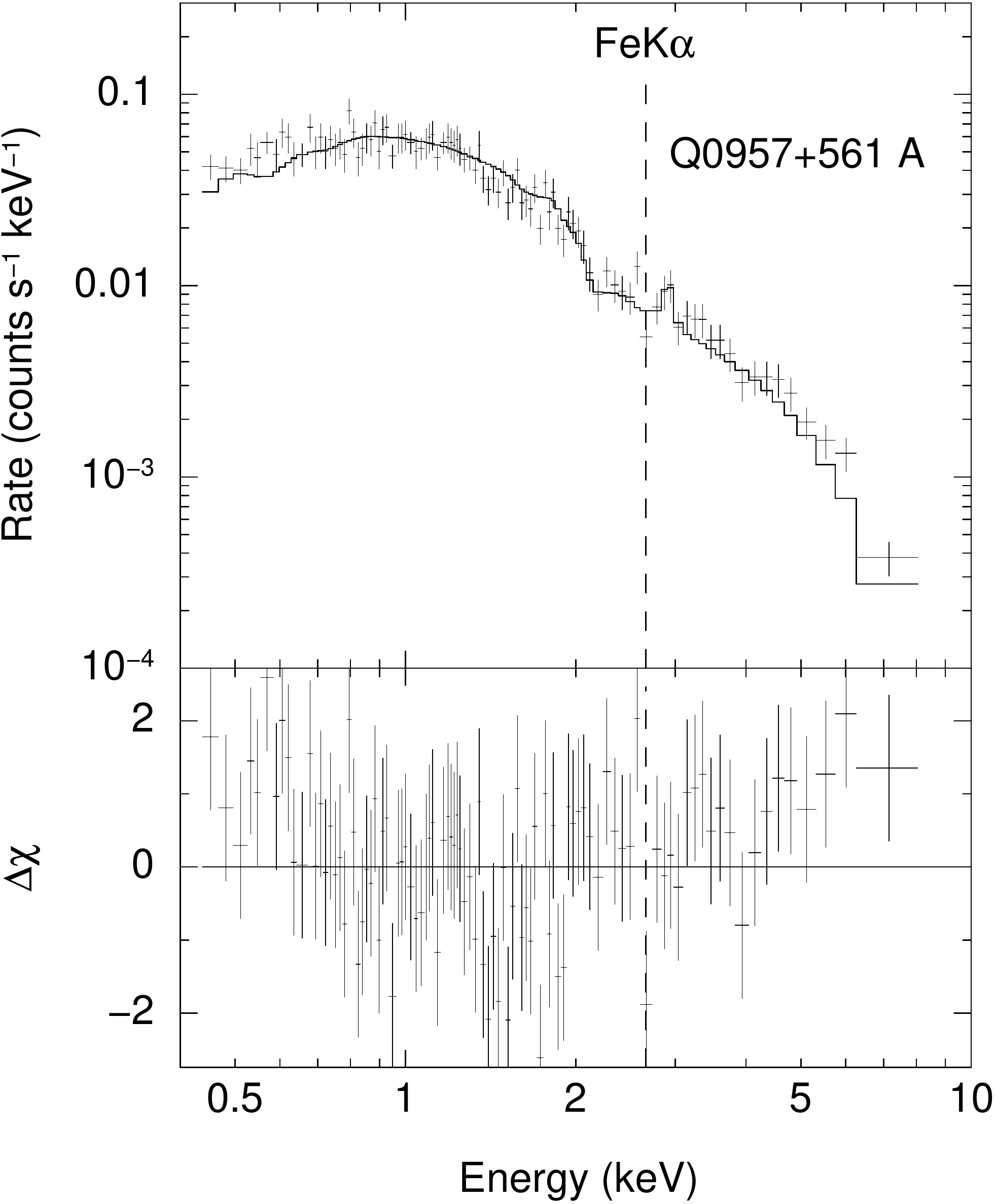} 
	\includegraphics[scale=0.44]{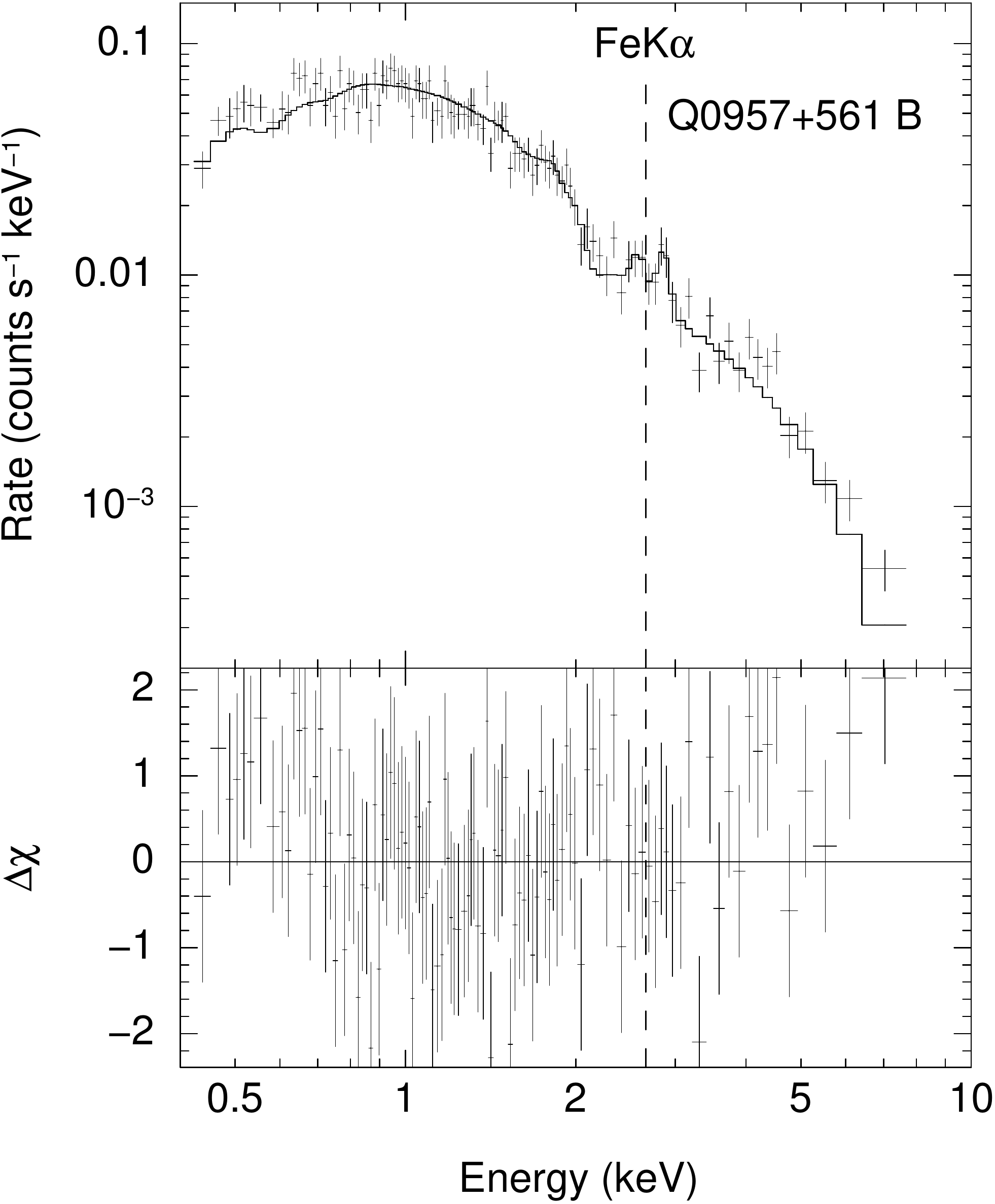}
	\caption{Stacked spectra of Q\,0957+561 and spectral fits. The sub-panels show the statistical residuals.}
	\label{specfit0957}
\end{figure}

We first extracted the spectra of individual images with CIAO, using circles of radii $\sim0\farcs8$ centred on the positions from the PSF fits for each observation. To estimate the background, we used the method given in \cite{chen2012}, which tries to account for the background contamination from the adjacent images of the lens. We then acquired the stacked spectra of individual images by combining all epochs and we used \textit{XSPEC} \citep{arnaud1996} to analyse the spectra. We modelled the spectra using a power law modified by Galactic absorption and lens galaxy absorption. We also added Gaussian emission lines to the models. During the spectral fitting which was performed within the energy range of 0.4--8 keV, we allowed the power law index ($\Gamma$) to vary, assumed the same Galactic absorption for all images fixed at the value calculated by \cite{dickey1990}, and set the $N_H$ of the lens galaxy free so that the absorption from the lens galaxy could vary independently. After fitting all the spectra, we calculated the absorbed to unabsorbed flux ratio ($f_{\text{abs}}/f_{\text{unabs}}$) for each image which we used for acquiring the absorption corrected count rates. We give these absorption corrected count rates in Tables \ref{corfluxq}, \ref{corfluxmg} and \ref{corfluxb}. The results of the spectral fit are presented in Figures \ref{specfit0414}, \ref{specfit1422}, and \ref{specfit0957} while the resulting parameters are listed in Tables \ref{tabspec_mg}, \ref{tabspec_b}, and \ref{tabspec_q}. Finally, we obtained the flux variations which are free from Galactic and lens galaxy absorptions. 

\begin{table}[htbp]
  \centering
  \caption{Absorption corrected count rates for Q\,0957+561.}
    \begin{tabular}{ccccccccc}
    \hline
    \hline
    Obs ID & Date  & Exp   & A$_{\text{full}}$ & A$_{\text{soft}}$ & A$_{\text{hard}}$ & B$_{\text{full}}$ & B$_{\text{soft}}$ & B$_{\text{hard}}$ \bigstrut\\
    \hline
    362   & 16 Apr 2000 & 47.662 & $240.1_{-14.0}^{+13.7}$ & $154.5_{-9.1}^{+8.9}$ & $85.4_{-5.1}^{+5.0}$ & $180.2_{-9.5}^{+9.6}$ & $114.4_{-6.1}^{+6.2}$ & $65.6_{-3.7}^{+3.7}$ \bigstrut[t]\\
    12076 & 17 Jan 2010 & 2.990 & $160.4_{-15.6}^{+13.1}$ & $69.0_{-7.4}^{+7.8}$ & $77.7_{-9.3}^{+12.5}$ & $97.8_{-8.6}^{+8.5}$ & $41.0_{-5.8}^{+5.9}$ & $48.4_{-5.6}^{+7.0}$ \\
    12077 & 1 Feb 2010 & 3.113 & $78.3_{-12.2}^{+9.5}$ & $37.5_{-4.6}^{+4.8}$ & $40.6_{-5.6}^{+5.8}$ & $105.0_{-13.0}^{+13.2}$ & $52.5_{-6.5}^{+6.8}$ & $60.6_{-9.0}^{+7.0}$ \\
    12078 & 18 Feb 2010 & 3.108 & $97.1_{-8.2}^{+8.1}$ & $42.0_{-5.9}^{+5.0}$ & $54.3_{-6.8}^{+5.6}$ & $110.1_{-9.2}^{+9.0}$ & $49.8_{-6.5}^{+6.5}$ & $55.5_{-6.9}^{+6.3}$ \\
    12079 & 3 Mar 2010 & 3.077 & $86.4_{-9.4}^{+18.0}$ & $44.1_{-6.4}^{+5.7}$ & $50.8_{-7.4}^{+5.9}$ & $87.6_{-8.2}^{+10.1}$ & $41.8_{-6.0}^{+5.6}$ & $50.9_{-7.4}^{+6.2}$ \\
    12080 & 15 Mar 2010 & 3.055 & $89.5_{-11.5}^{+10.5}$ & $51.8_{-12.1}^{+5.5}$ & $47.9_{-6.8}^{+5.6}$ & $96.3_{-15.6}^{+9.7}$ & $43.9_{-10.5}^{+5.5}$ & $53.3_{-6.2}^{+6.6}$ \\
    12081 & 29 Mar 2010 & 3.097 & $80.9_{-9.9}^{+9.3}$ & $39.6_{-6.1}^{+6.6}$ & $42.1_{-5.2}^{+5.6}$ & $61.2_{-7.3}^{+6.6}$ & $30.3_{-4.5}^{+4.7}$ & $26.7_{-4.8}^{+4.3}$ \\
    12082 & 13 Apr 2010 & 3.109 & $82.3_{-19.2}^{+7.6}$ & $37.7_{-10.3}^{+4.9}$ & $42.1_{-7.8}^{+5.5}$ & $71.7_{-6.4}^{+6.5}$ & $34.8_{-9.3}^{+4.1}$ & $35.8_{-8.3}^{+4.6}$ \\
    12083 & 27 Apr 2010 & 3.079 & $80.4_{-15.0}^{+7.8}$ & $38.7_{-5.1}^{+5.2}$ & $42.2_{-5.2}^{+4.7}$ & $77.6_{-13.1}^{+7.6}$ & $38.4_{-4.6}^{+4.6}$ & $41.3_{-5.6}^{+4.9}$ \\
    12084 & 15 May 2010 & 3.109 & $62.4_{-6.5}^{+6.6}$ & $24.3_{-3.2}^{+17.6}$ & $32.8_{-3.7}^{+4.8}$ & $153.9_{-12.2}^{+11.6}$ & $61.8_{-11.1}^{+11.5}$ & $75.8_{-6.7}^{+8.6}$ \\
    12085 & 25 May 2010 & 2.992 & $84.9_{-11.4}^{+8.3}$ & $38.5_{-4.6}^{+5.2}$ & $35.2_{-3.8}^{+5.9}$ & $128.1_{-12.4}^{+12.1}$ & $66.4_{-8.9}^{+7.5}$ & $57.0_{-6.0}^{+9.8}$ \\
    12086 & 10 Jun 2010 & 2.992 & $91.2_{-8.6}^{+8.9}$ & $44.0_{-5.6}^{+5.4}$ & $48.1_{-6.3}^{+6.0}$ & $134.3_{-12.2}^{+11.1}$ & $68.6_{-6.6}^{+6.5}$ & $66.9_{-7.4}^{+7.1}$ \\
    12087 & 23 Jun 2010 & 2.992 & $90.9_{-25.2}^{+19.6}$ & $45.8_{-7.7}^{+5.0}$ & $47.1_{-7.0}^{+6.4}$ & $132.2_{-41.3}^{+10.3}$ & $58.2_{-7.0}^{+6.7}$ & $64.3_{-8.6}^{+9.3}$ \bigstrut[b]\\
    \hline
    \end{tabular}%
  \label{corfluxq}%
\tablecomments{Count rates are in units of $10^{-3}s^{-1}$. Exposure time is given under ``Exp" in units of $10^3 s$.} 
\end{table}%

\begin{sidewaystable}[htbp]
  \centering
  \caption{Absorption corrected count rates for MG\,J0414+0534.}
    \begin{tabular}{ccccccccccccccc}
    \hline
    \hline
    Obs ID & Date  & Exp   & A$_{\text{full}}$ & A$_{\text{soft}}$ & A$_{\text{hard}}$ & B$_{\text{full}}$ & B$_{\text{soft}}$ & B$_{\text{hard}}$ & C$_{\text{full}}$ & C$_{\text{soft}}$ & C$_{\text{hard}}$ & D$_{\text{full}}$ & D$_{\text{soft}}$ & D$_{\text{hard}}$ \bigstrut\\
    \hline
    417   & 13 Jan 2000 & 6.579 & $30.0_{-3.5}^{+3.5}$ & $9.9_{-2.1}^{+2.1}$ & $18.6_{-2.5}^{+3.3}$ & $30.8_{-3.1}^{+3.7}$ & $8.1_{-1.6}^{+1.8}$ & $19.6_{-3.2}^{+4.0}$ & $15.3_{-2.0}^{+2.2}$ & $5.4_{-1.2}^{+1.3}$ & $8.1_{-1.2}^{+2.8}$ & $3.7_{-0.9}^{+1.0}$ & $1.1_{-0.4}^{+0.5}$ & $2.7_{-0.7}^{+1.0}$ \bigstrut[t]\\
    418   & 2 Apr 2000 & 7.440 & $22.9_{-3.0}^{+3.5}$ & $7.0_{-1.6}^{+1.5}$ & $15.6_{-2.8}^{+2.8}$ & $35.8_{-3.9}^{+3.3}$ & $10.6_{-1.5}^{+2.1}$ & $24.4_{-7.5}^{+3.0}$ & $14.2_{-1.6}^{+1.7}$ & $5.1_{-1.2}^{+1.1}$ & $9.5_{-1.8}^{+1.6}$ & $9.3_{-1.7}^{+1.8}$ & $2.8_{-0.7}^{+1.1}$ & $6.0_{-1.3}^{+1.4}$ \\
    421   & 16 Aug 2000 & 7.251 & $21.4_{-2.9}^{+2.9}$ & $5.7_{-1.2}^{+1.5}$ & $15.3_{-2.3}^{+2.4}$ & $37.5_{-3.9}^{+4.3}$ & $11.8_{-1.7}^{+2.0}$ & $26.3_{-5.0}^{+3.1}$ & $16.4_{-1.9}^{+2.3}$ & $5.8_{-1.0}^{+1.2}$ & $9.9_{-1.3}^{+1.4}$ & $7.6_{-1.5}^{+1.6}$ & $2.7_{-0.8}^{+0.9}$ & $5.0_{-1.1}^{+1.2}$ \\
    422   & 16 Nov 2000 & 7.504 & $24.8_{-3.5}^{+4.8}$ & $7.5_{-1.4}^{+1.7}$ & $16.6_{-2.4}^{+2.8}$ & $39.4_{-5.0}^{+4.7}$ & $12.5_{-5.7}^{+1.8}$ & $27.2_{-4.3}^{+4.5}$ & $13.1_{-1.7}^{+2.0}$ & $4.9_{-0.9}^{+1.0}$ & $9.2_{-1.5}^{+1.7}$ & $9.6_{-1.7}^{+1.9}$ & $2.2_{-0.6}^{+0.7}$ & $7.1_{-1.4}^{+1.7}$ \\
    1628  & 5 Feb 2001 & 9.024 & $25.4_{-4.3}^{+3.2}$ & $6.3_{-1.3}^{+1.5}$ & $21.5_{-2.7}^{+2.7}$ & $40.9_{-12.6}^{+3.7}$ & $13.2_{-1.7}^{+1.7}$ & $26.4_{-2.7}^{+2.9}$ & $18.1_{-1.9}^{+2.1}$ & $6.9_{-1.0}^{+1.1}$ & $11.9_{-1.4}^{+1.5}$ & $6.8_{-1.2}^{+1.3}$ & $2.1_{-0.6}^{+0.7}$ & $4.8_{-1.0}^{+1.1}$ \\
    3395  & 9 Nov 2001 & 28.416 & $23.5_{-4.8}^{+2.3}$ & $6.4_{-0.8}^{+0.9}$ & $17.2_{-1.7}^{+1.7}$ & $26.1_{-2.4}^{+1.7}$ & $8.5_{-0.9}^{+1.0}$ & $18.9_{-1.4}^{+1.6}$ & $14.6_{-3.2}^{+1.2}$ & $4.3_{-0.5}^{+0.5}$ & $9.7_{-0.9}^{+0.9}$ & $7.2_{-2.0}^{+1.0}$ & $2.0_{-0.4}^{+0.4}$ & $5.5_{-0.8}^{+0.9}$ \\
    3419  & 8 Jan 2002 & 96.663 & $23.2_{-1.8}^{+1.8}$ & $7.0_{-0.7}^{+0.7}$ & $18.6_{-1.5}^{+1.5}$ & $33.4_{-1.6}^{+1.7}$ & $9.1_{-0.6}^{+0.6}$ & $21.4_{-1.2}^{+1.2}$ & $15.1_{-0.8}^{+0.8}$ & $4.9_{-0.3}^{+0.3}$ & $10.9_{-0.7}^{+0.7}$ & $7.3_{-0.9}^{+0.9}$ & $2.3_{-0.3}^{+0.3}$ & $4.7_{-0.6}^{+0.6}$ \\
    12800 & 15 Oct 2011 & 29.677 & $14.7_{-1.5}^{+1.6}$ & $3.7_{-0.6}^{+0.7}$ & $11.8_{-1.3}^{+1.3}$ & $20.7_{-1.7}^{+1.6}$ & $4.8_{-0.6}^{+0.8}$ & $14.4_{-1.2}^{+1.3}$ & $7.2_{-0.6}^{+0.6}$ & $1.8_{-0.4}^{+0.3}$ & $5.9_{-0.6}^{+0.6}$ & $4.0_{-0.6}^{+0.7}$ & $0.6_{-0.2}^{+0.2}$ & $3.3_{-0.5}^{+0.6}$ \bigstrut[b]\\
    \hline
    \end{tabular}
  \label{corfluxmg}
  \tablecomments{Count rates are in units of $10^{-3}s^{-1}$. Exposure time is given under ``Exp" in units of $10^3 s$.}
  \vspace{50px}

    \centering
  \caption{Absorption corrected count rates for B\,1422+231.}
    \begin{tabular}{ccccccccccccccc}
    \hline
    \hline
    Obs ID & Date  & Exp   & A$_{\text{full}}$ & A$_{\text{soft}}$ & A$_{\text{hard}}$ & B$_{\text{full}}$ & B$_{\text{soft}}$ & B$_{\text{hard}}$ & C$_{\text{full}}$ & C$_{\text{soft}}$ & C$_{\text{hard}}$ & D$_{\text{full}}$ & D$_{\text{soft}}$ & D$_{\text{hard}}$ \bigstrut\\
    \hline
    367   & 1 Jun 2000 & 28.464 & $41.0_{-2.8}^{+2.7}$ & $16.1_{-1.5}^{+1.6}$ & $22.5_{-1.6}^{+1.7}$ & $54.6_{-3.6}^{+4.6}$ & $20.9_{-1.9}^{+2.4}$ & $29.6_{-2.5}^{+2.9}$ & $28.0_{-1.3}^{+1.4}$ & $14.3_{-1.1}^{+1.2}$ & $16.4_{-1.1}^{+1.2}$ & $3.3_{-0.4}^{+0.4}$ & $1.6_{-0.3}^{+0.3}$ & $1.6_{-0.3}^{+0.3}$ \bigstrut[t]\\
    1631  & 21 May 2001 & 10.652 & $41.6_{-4.0}^{+3.3}$ & $15.8_{-1.9}^{+2.3}$ & $20.7_{-2.2}^{+2.4}$ & $54.5_{-5.4}^{+5.4}$ & $21.7_{-2.7}^{+5.0}$ & $30.2_{-3.4}^{+3.7}$ & $31.6_{-1.9}^{+2.0}$ & $18.2_{-3.5}^{+1.9}$ & $17.7_{-1.8}^{+1.8}$ & $2.9_{-0.6}^{+0.6}$ & $1.5_{-0.4}^{+0.6}$ & $1.6_{-0.4}^{+0.5}$ \\
    4939  & 1 Dec 2004 & 47.730 & $37.9_{-2.2}^{+2.2}$ & $14.9_{-1.2}^{+1.3}$ & $21.1_{-1.4}^{+1.4}$ & $45.9_{-3.2}^{+3.9}$ & $19.2_{-1.6}^{+3.3}$ & $27.4_{-1.9}^{+2.4}$ & $28.6_{-1.1}^{+1.2}$ & $12.4_{-1.4}^{+0.7}$ & $16.4_{-0.8}^{+0.9}$ & $3.1_{-0.3}^{+0.3}$ & $1.4_{-0.2}^{+0.6}$ & $1.8_{-0.2}^{+0.3}$ \\
    12801 & 24 Nov 2011 & 29.587 & $24.5_{-1.6}^{+1.6}$ & $7.2_{-0.6}^{+0.8}$ & $15.0_{-1.2}^{+4.3}$ & $38.1_{-2.7}^{+3.5}$ & $11.6_{-1.1}^{+1.6}$ & $24.2_{-2.3}^{+2.7}$ & $34.8_{-1.6}^{+1.7}$ & $14.6_{-1.2}^{+1.0}$ & $22.6_{-5.1}^{+1.4}$ & $2.9_{-0.4}^{+0.4}$ & $1.1_{-0.2}^{+0.2}$ & $2.0_{-0.3}^{+1.0}$ \bigstrut[b]\\
    \hline
    \end{tabular}%
  \label{corfluxb}%
	\tablecomments{Count rates are in units of $10^{-3}s^{-1}$. Exposure time is given under ``Exp" in units of $10^3 s$.}
\end{sidewaystable}

\begin{table}[htbp]
\resizebox{\textwidth}{!}{
\begin{threeparttable}
  \centering
  \caption{Spectral Fit Results For MG\,J0414+0534.}
    \begin{tabular}{ccccccccc}
    \toprule
    \toprule
    Image & $\Gamma$ & $N_H$ & $E_{\text{line}}$ (keV) & $\sigma_{\text{line}}$ (keV) & EW (keV) & Flux  & $\chi_\nu^2$ & $P(\chi_\nu^2)$ \\
          &       & $(\times 10^{22}\,cm^{-2})$ &       &       &       & $(\times 10^{-13}\,erg\,cm^{-2}\,s^{-1})$ &       &  \\
    \midrule
    A     & $1.68_{-0.09}^{+0.10}$ & $0.69_{-0.12}^{+0.14}$ & $6.52_{-0.28}^{+0.40}$ & $0.12^\ast$ & $0.20_{-0.02}^{+0.07}$ & $1.54_{-0.08}^{+0.09}$ & 0.93  & 0.64 \\
    B     & $1.67_{-0.06}^{+0.06}$ & $0.95_{-0.10}^{+0.11}$ & $\ldots$ & $\ldots$ & $\ldots$ & $5.10_{-0.17}^{+0.17}$ & 1.27  & 0.03 \\
    C     & $1.66_{-0.06}^{+0.06}$ & $0.98_{-0.09}^{+0.10}$ & $\ldots$ & $\ldots$ & $\ldots$ & $5.81_{-0.18}^{+0.18}$ & 1.24  & 0.02 \\
    D     & $1.74_{-0.17}^{+0.19}$ & $1.02_{-0.28}^{+0.27}$ & $\ldots$ & $\ldots$ & $\ldots$ & $0.80_{-0.07}^{+0.07}$ & 1.10  & 0.32 \\
    \bottomrule
    \end{tabular}%
  \label{tabspec_mg}
\begin{tablenotes}[para, flushleft]
	\item \textbf{Notes:} Reduced $\chi^2$ is defined by $\chi^2_\nu=\chi^2/\nu$ where $\nu$ is the degree of freedom. Errors are derived at 68\% confidence level. The last column gives the probability of exceeding $\chi^2$ for $\nu$ degrees of freedom. Parameters marked with an asterisk are unconstrained. 
\end{tablenotes}
\end{threeparttable}}

\vspace{10px}


  \centering
  \caption{Spectral Fit Results For B\,1422+231}
    \begin{tabular}{cccccc}
    \hline
    \hline
    Image & $\Gamma$ & $N_H$ & Flux  & $\chi_\nu^2$ & $P(\chi_\nu^2)$ \bigstrut[t]\\
          &       & $(\times 10^{22}\,cm^{-2})$ & $(\times 10^{-13}\,erg\,cm^{-2}\,s^{-1})$ &       &  \bigstrut[b]\\
    \hline
    A     & $1.51_{-0.04}^{+0.05}$ & $0.00_{-0.00}^{+0.01}$ & $4.23_{-0.15}^{+0.16}$ & 1.25  & 0.02 \bigstrut[t]\\
    B     & $1.57_{-0.05}^{+0.05}$ & $0.00_{-0.00}^{+0.02}$ & $3.08_{-0.13}^{+0.13}$ & 1.29  & 0.02 \\
    C     & $1.55_{-0.14}^{+0.14}$ & $0.00_{-0.00}^{+0.02}$ & $0.49_{-0.06}^{+0.06}$ & 1.31  & 0.20 \\
    D     & $1.56_{-0.04}^{+0.04}$ & $0.00_{-0.00}^{+0.01}$ & $3.63_{-0.14}^{+0.14}$ & 1.22  & 0.05 \bigstrut[b]\\
    \hline
    \end{tabular}
  \label{tabspec_b}
  \vspace{15px}
\end{table}

\begin{table}[htbp]
  \centering
  \caption{Spectral Fit Results For Q\,0957+561}
  \resizebox{\textwidth}{!}{
    \begin{tabular}{cccccccccccc}
    \toprule
    \toprule
    Image & $\Gamma$ & $N_H$ & $E_{\text{line1}}$ & $\sigma_{\text{line1}}$ & EW Line1 & $E_{\text{line2}}$ (keV) & $\sigma_{\text{line2}}$ & EW Line2 & Flux  & $\chi_\nu^2$ & $P(\chi_\nu^2)$ \\
          &       & $(\times 10^{22}\,cm^{-2})$ & (keV) & (keV) & (keV) & (keV) & (keV) & (keV) & $(\times 10^{-13}\,erg\,cm^{-2}\,s^{-1})$ &       &  \\
    \midrule
    A     & $1.99_{-0.06}^{+0.07}$ & $0.00_{-0.00}^{+0.01}$ & $7.02_{-0.16}^{+0.15}$ & $0.10^\ast$ & $0.33_{-0.04}^{+0.14}$ & $\ldots$ & $\ldots$ & $\ldots$ & $6.30_{-0.27}^{+0.26}$ & 1.15  & 0.15 \\
    B     & $2.01_{-0.06}^{+0.06}$ & $0.00_{-0.00}^{+0.01}$ & $6.88_{-0.10}^{+0.09}$ & $<0.14$ & $0.31_{-0.07}^{+0.15}$ & $6.23_{-0.16}^{+0.16}$ & $0.10^\ast$ & $0.29_{-0.05}^{+0.12}$ & $7.27_{-0.26}^{+0.27}$ & 1.07  & 0.30 \\
    \bottomrule
    \end{tabular}}
  \label{tabspec_q}
\end{table}

\subsection{Emission Lines}

We tentatively detected FeK$\alpha$ fluorescence line in image A of MG\,J0414+0534, confirming the earlier detection by \cite{chartas2002}, and in both images of Q\,0957+561, but not in B\,1422+231. As can be seen from Tables \ref{tabspec_mg} and \ref{tabspec_q}, the rest frame energies of the detected FeK$\alpha$ lines are consistent with the neutral FeK$\alpha$ emission at 6.4 keV. Shifts in the line energy are seen in both Q\,0957+561 A and B. We also found that adding two lines instead of one in image B of Q\,0957+561 significantly improved the fit. In this case, we measure a redshifted line at 6.23 keV and a blueshifted line at 6.88 keV. Such FeK$\alpha$ line shifts have previously been detected in a sample of radio-quiet lensed quasars \citep{chen2012, chartas2017}.

\begin{figure}
	\centering
	\includegraphics[scale=0.75]{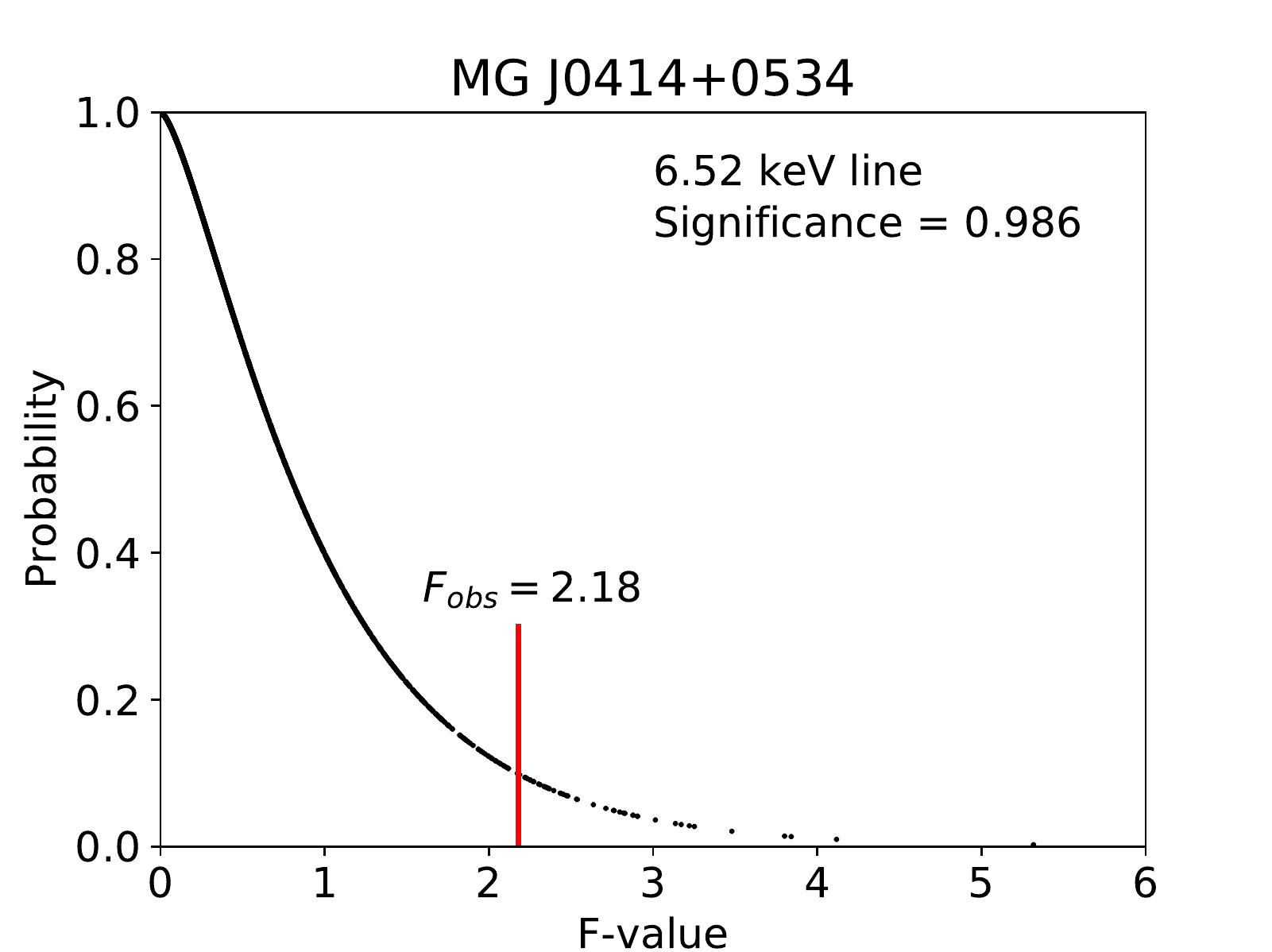}
	\caption{$F$-statistic distribution derived from Monte Carlo simulations for image A of MG\,J0414+0534.}
	\label{fsimmg}
\end{figure}

\begin{figure}
	\centering
	\includegraphics[scale=0.75]{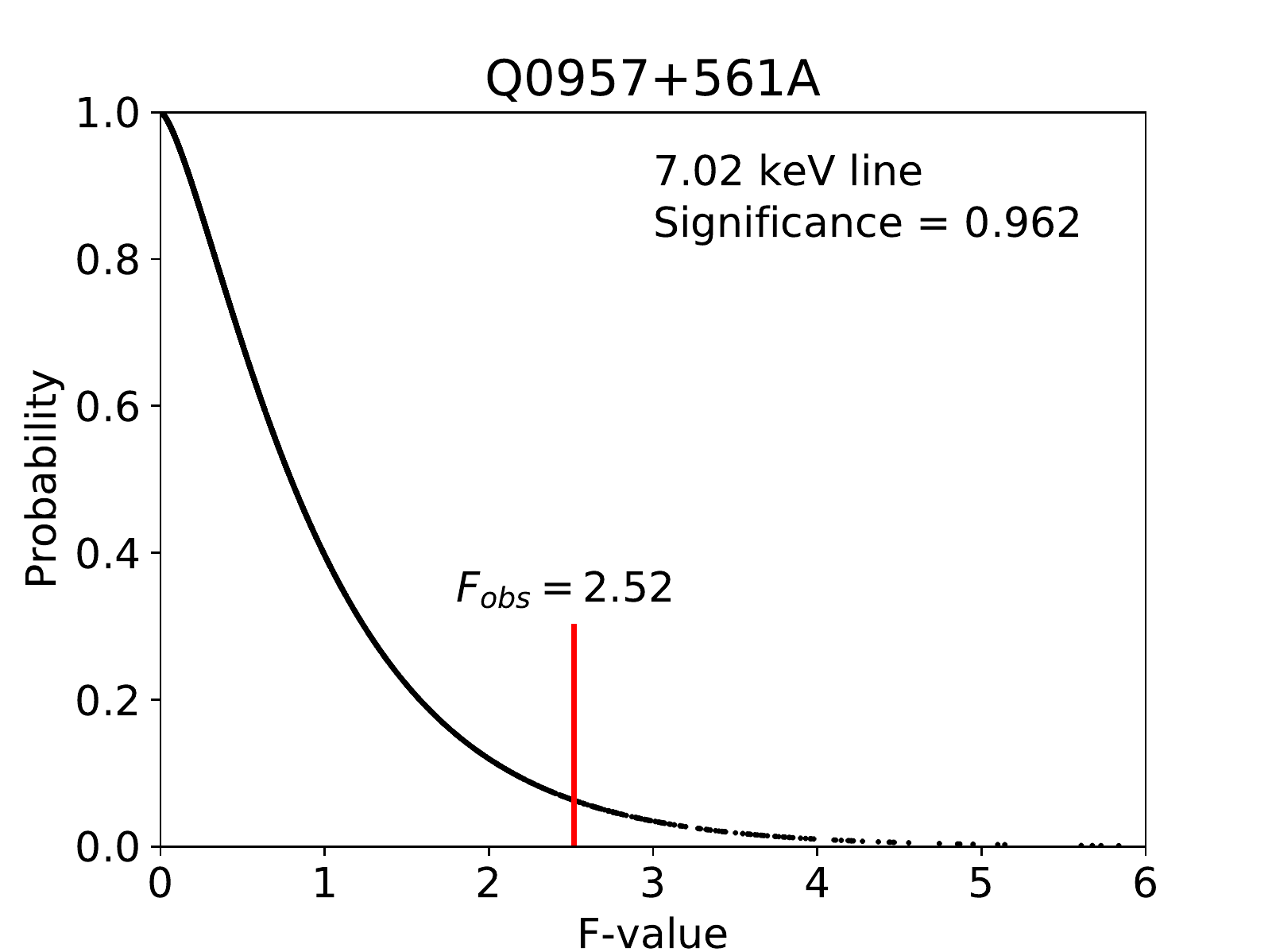}
	\caption{$F$-statistic distribution derived from Monte Carlo simulations for image A of Q\,0957+561.}
	\label{fsimqa}
\end{figure}

\begin{figure}
	\centering
	\includegraphics[scale=0.55]{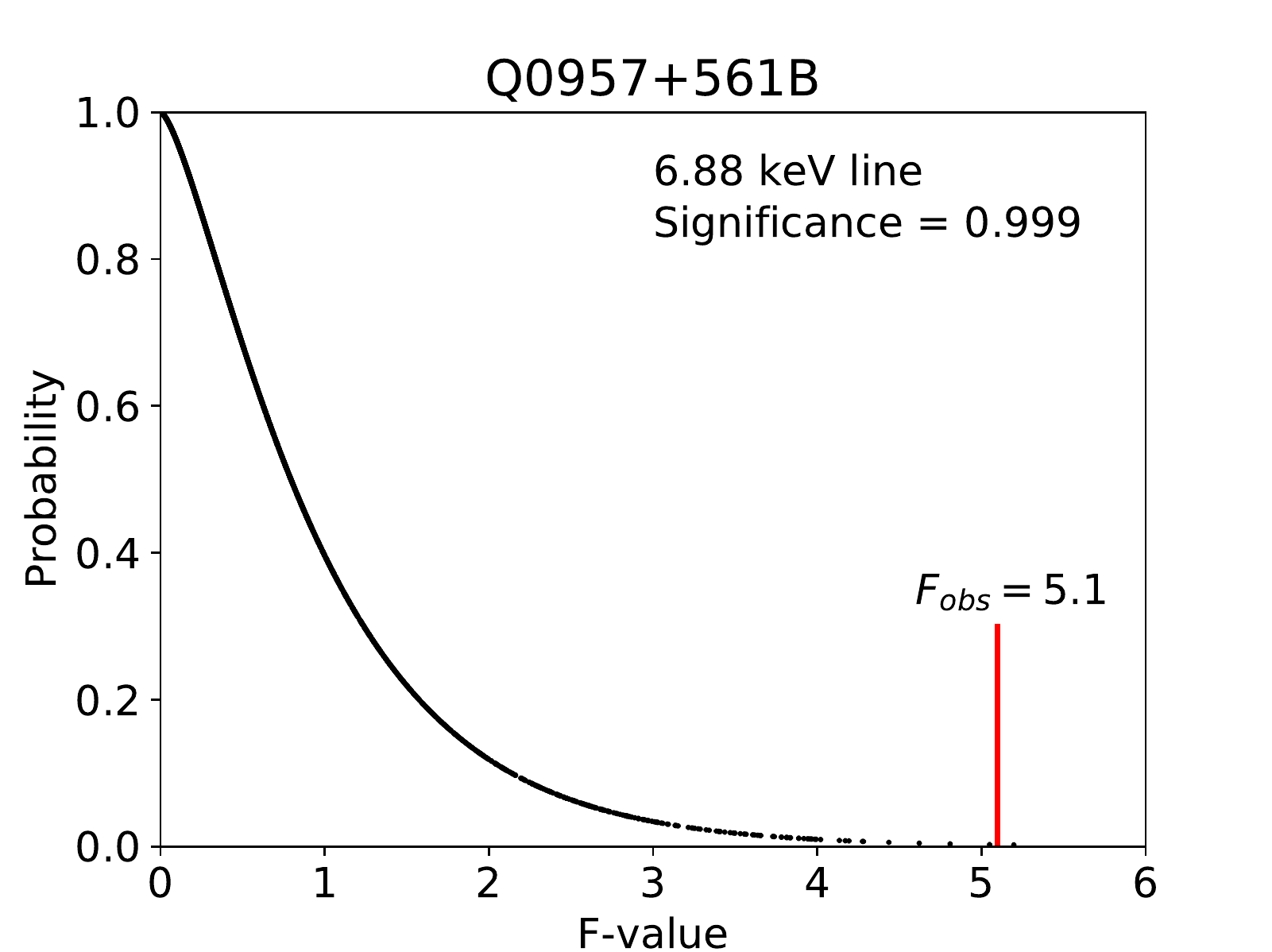}
	\includegraphics[scale=0.55]{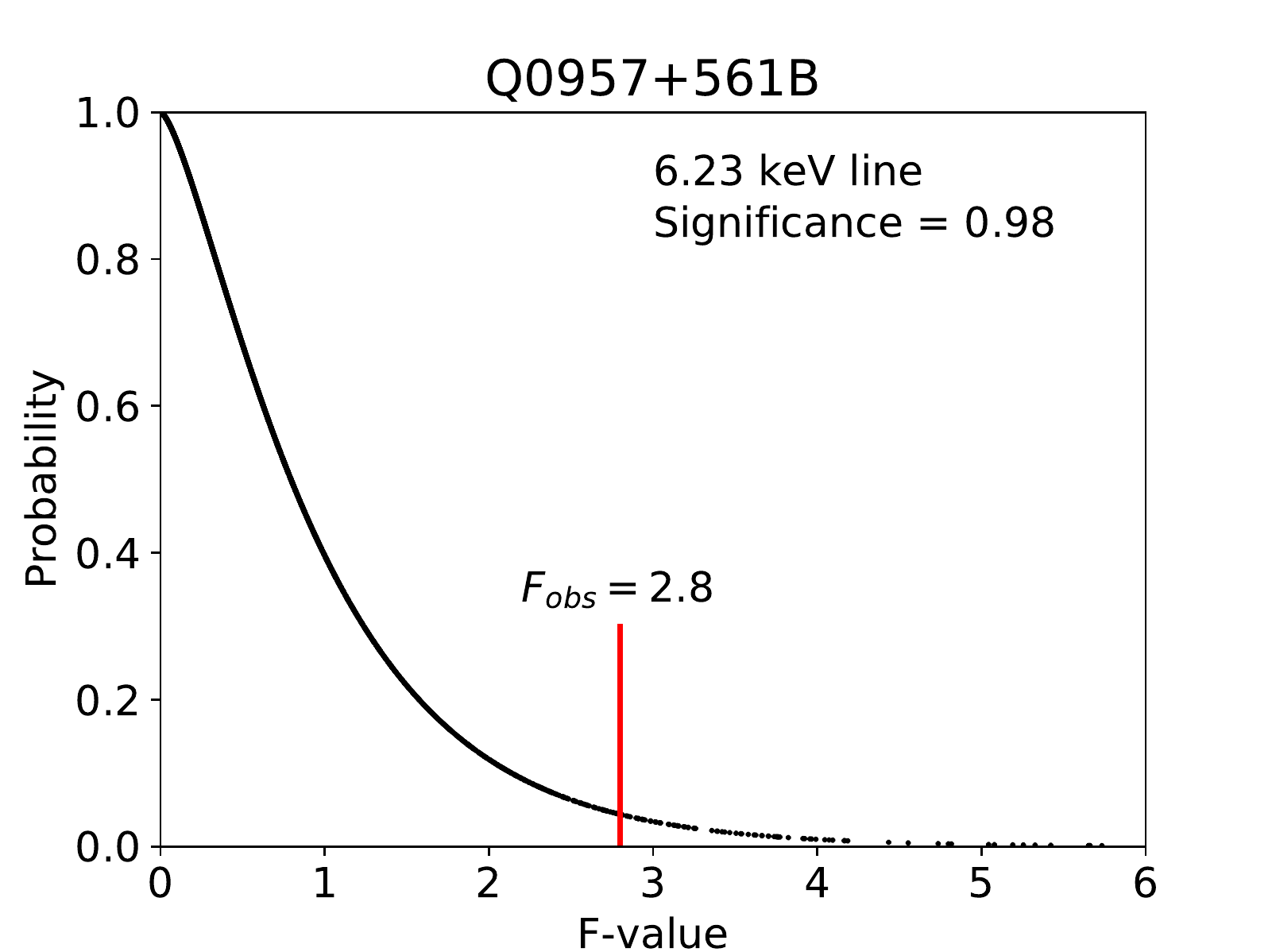}
	\caption{$F$-statistic distributions derived from Monte Carlo simulations for image B of Q\,0957+561.}
	\label{fsimqb}
\end{figure}

To calculate the statistical significance of the detected emission features, we used a Monte Carlo simulation approach proposed by \cite{protas2002}. From this, we determined the distribution of the $F$-statistic between the null model (absorbed power law) with no emission lines and the alternative model (absorbed power law including one or more Gaussian emission lines) for 5000 spectra simulated from the null model with XSPEC. Each simulated spectrum was binned the same as the actual spectrum, and fitted with the null model, then fitted again with the alternative model. After these fits for two different models, $F$-test was performed for each simulation, and finally, the statistical significance value was calculated by comparing the $F$-test values from simulations and the ones from real data ($F_{\text{obs}}$). Additionally, analytical significance was obtained from the probability corresponding to $F_{\text{obs}}$, i.e. result of $F$-test applied to the data. The results of the simulations are shown in Figures \ref{fsimmg}, \ref{fsimqa}, and \ref{fsimqb}. The significance values are given in Table \ref{linesig}.

\begin{table}[htbp]
  \centering
  \caption{Significance of the Detected Lines}
    \begin{tabular}{ccccc}
    \toprule \toprule
    Lens  & Image & $E_{\text{line}}$ (keV) & Monte Carlo & Analytical \\
          &       &       & Significance & Significance \\
    \midrule
    MG\,J0414+0534 & A     & $6.52_{-0.28}^{+0.40}$ & 98.61\% & 87.90\% \\
    Q\,0957+561 & A     & $7.02_{-0.16}^{+0.15}$ & 96.18\% & 91.41\% \\
    Q\,0957+562 & B     & $6.88_{-0.10}^{+0.09}$ & 99.92\% & 99.23\% \\
    Q\,0957+563 & B     & $6.23_{-0.16}^{+0.16}$ & 98.03\% & 93.46\% \\
    \bottomrule
    \end{tabular}%
  \label{linesig}
\end{table}

\section{Microlensing light curves}

\begin{figure}
	\centering
	\includegraphics[scale=0.75]{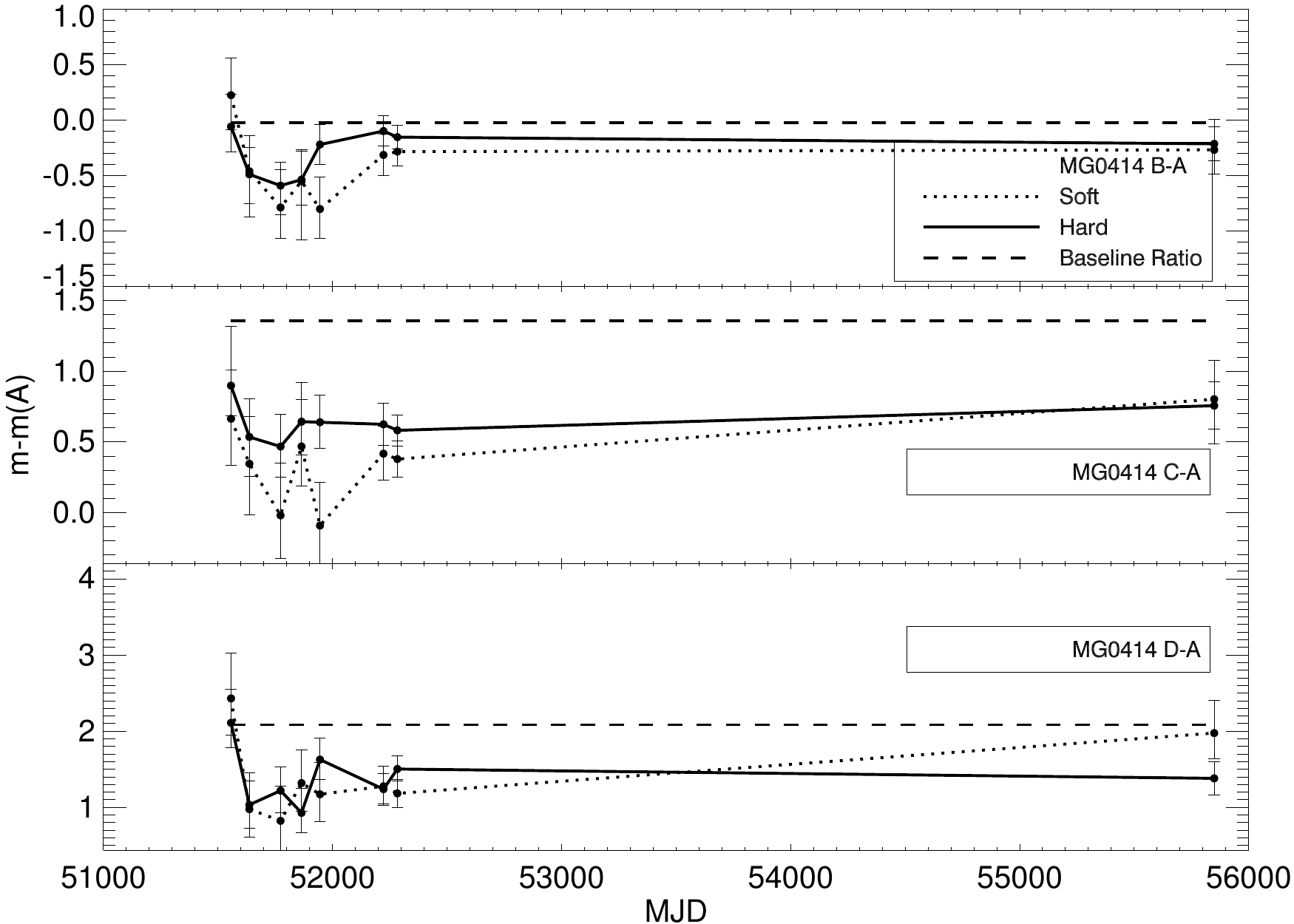}
	\caption{Microlensing light curves of MG\,J0414+0534 in magnitude scale.}
	\label{lc0414}
\end{figure}

\begin{figure}
	\centering
	\includegraphics[scale=0.75]{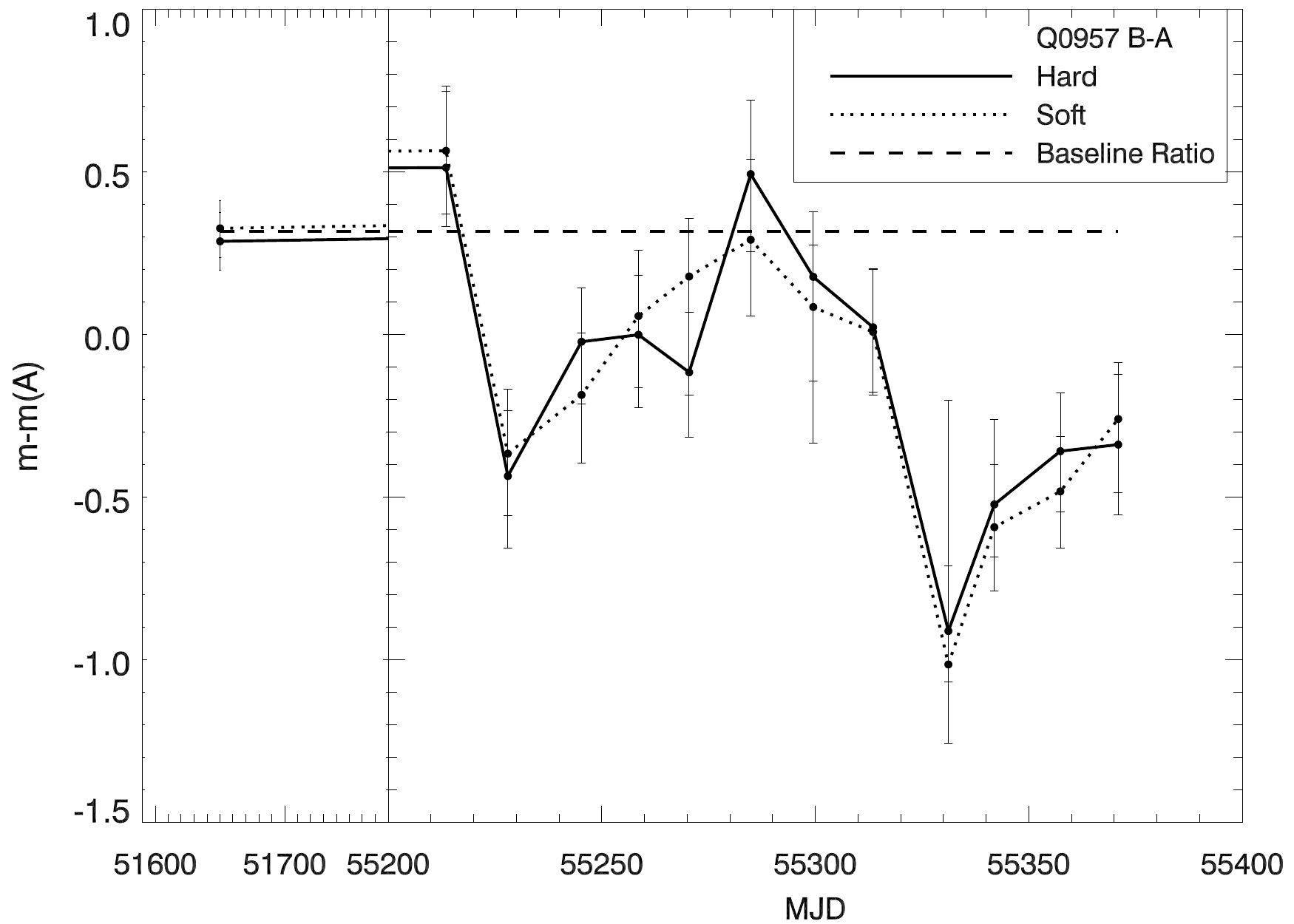}
	\caption{Microlensing light curves of Q\,0957+561 in magnitude scale.}
	\label{lc0957}
%
	\centering
	\includegraphics[scale=0.75]{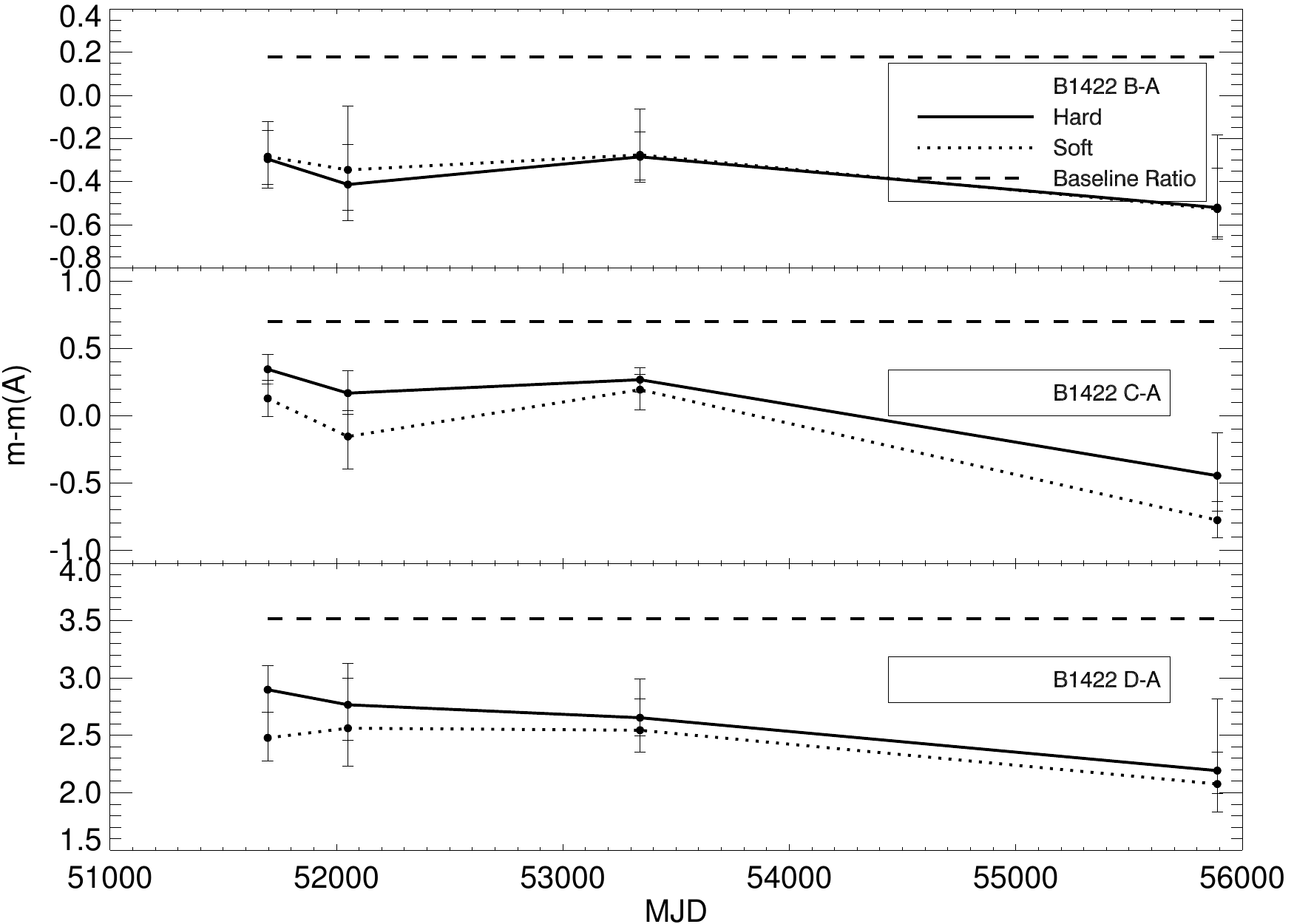}
	\caption{Microlensing light curves of B\,1422+231 in magnitude scale.}
	\label{lc1422}
\end{figure}

In this work, the microlensing light curves were measured based on the absorption corrected count rates given in Tables \ref{corfluxq} -- \ref{corfluxb}. Our aim was to analyse the differential microlensing light curves, the departure of the measured microlensed flux ratios from the intrinsic flux ratios \citep{guerras2017}. As for time-delay effects, as shown by \cite{sch2014}, the amplitude of source variability for luminous quasars in X-rays is small compared to both observational errors and microlensing amplitudes. This makes the source variability unlikely to contribute significantly to microlensing signal. We will explore this effect further by including quasar variability models in the microlensing analysis for long time-delay lenses (Cornachione et al. in preparation). We calculated the baseline flux ratios from the macrolensing models using the expression for magnification $\mu=1/ \lvert (1-\kappa^2)-\gamma^2 \rvert$ where $\kappa$ is the convergence (the dimensionless surface mass density of the lens galaxy) and $\gamma$ is the shear parameter which is responsible for the distortion of images. The $\kappa$ and $\gamma$ values for MG\,J0414+0534 and B\,1422+231 were taken from \cite{sch2014}, whereas the values for Q\,0957+561 were taken from \cite{media2009}. Baseline ratios are calculated with, for example between the A-B image pair, $-2.5\log(\mu_B/\mu_A)$. The microlensing light curves are shown in Figures \ref{lc0414}-\ref{lc1422}. Since the microlensing light curve depends only on flux ratios, the change of \textit{Chandra} effective area over time does not affect our microlensing light curves.

\begin{figure}
	\centering
	\includegraphics[scale=0.75]{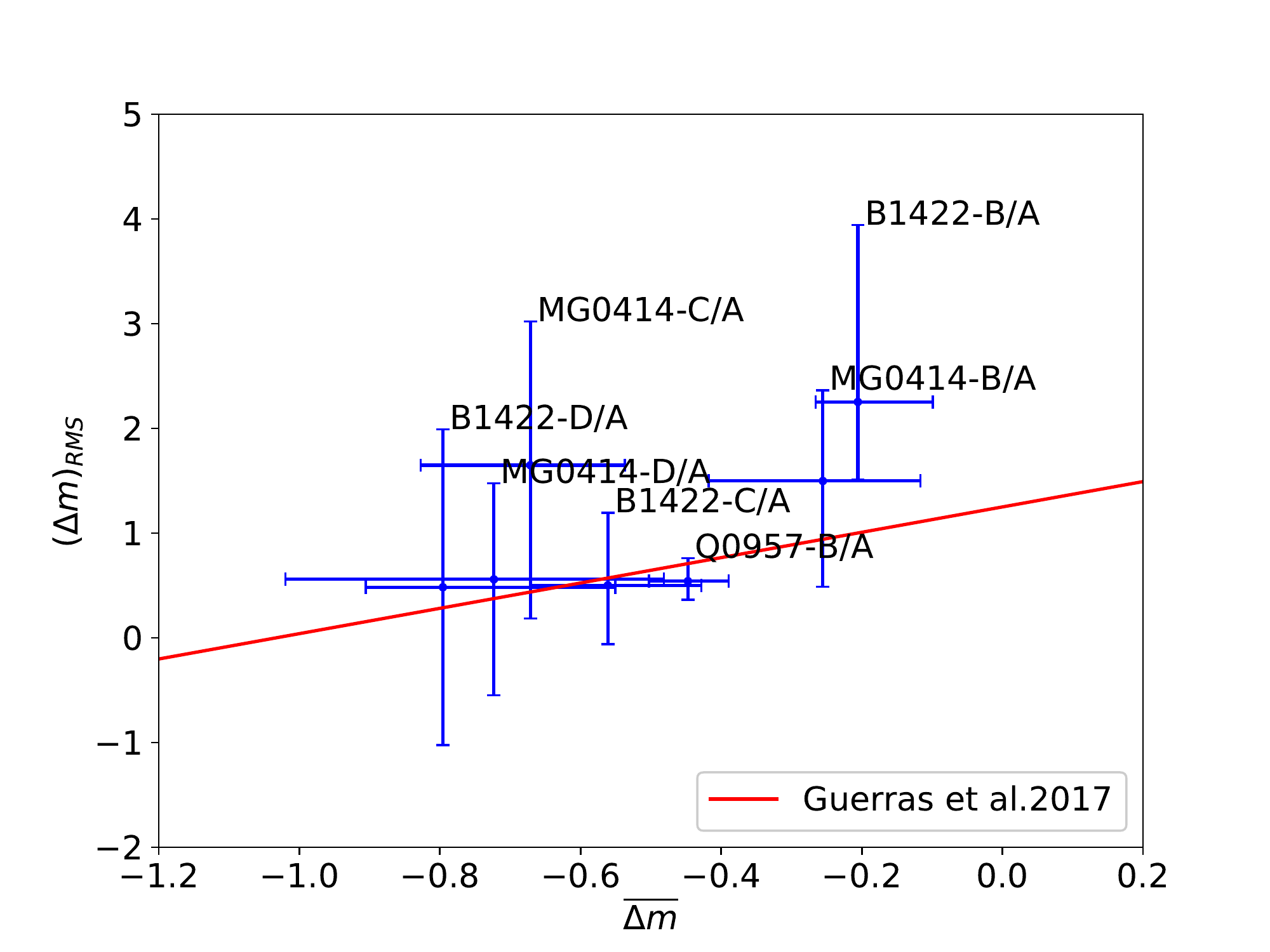}
	\includegraphics[scale=0.75]{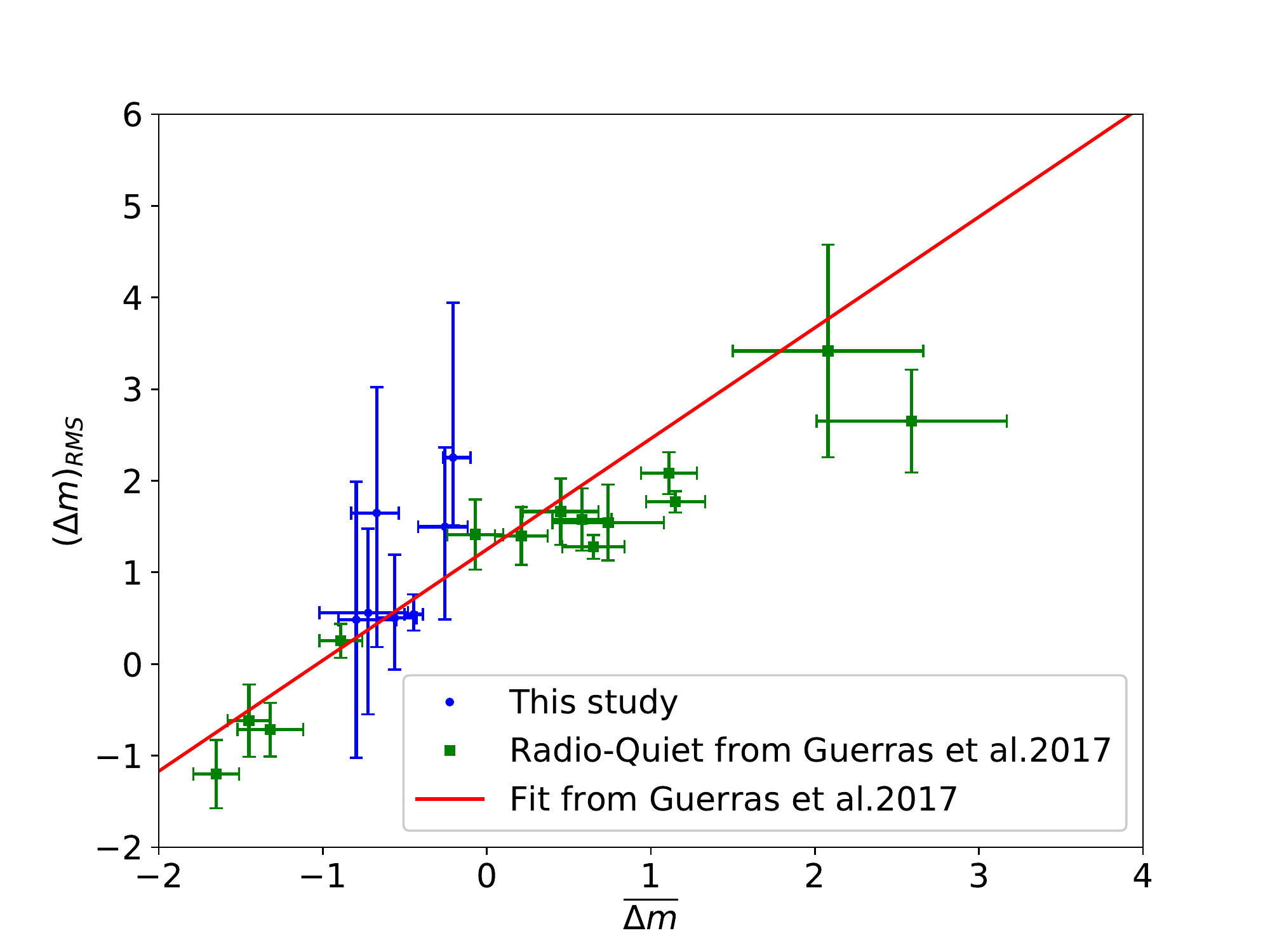}
	\caption{\textit{Upper panel:} RMS of microlensing variability in hard X-rays for different image pairs in our sample of radio-loud quasars. \textit{Lower panel:} Comparison with the radio-quiet sample from \cite{guerras2017}.}
	\label{rms}
\end{figure}

Continuing the notion from \cite{guerras2017} and \cite{guerras2018}, we also examine the root mean square (rms) of microlensing variability for our targets. Here, microlensing amplitudes $(\varphi)$ are the departures from the baseline ratio, and they can be calculated between images, e.g. A and B, at time $t_j$  from 
\begin{equation}
	\varphi_{AB}(t_j) = \frac{\varepsilon_{Bj}}{\varepsilon_{Aj}} = \frac{f_{Bj}}{f_{Aj}} \frac{\mu_A}{\mu_B}
	\label{xidef}
\end{equation}
where $f$ is the measured flux, $\mu$ is the macrolensing magnification, and $\varepsilon$ is the microlensing magnification. For each image pair, we calculate the mean microlensing amplitude $(\overline{\varphi})$ and its rms. Finally, we give the relation between these two parameters in Figure \ref{rms} in units of magnitudes where $\overline{\Delta m} = -2.5\log \overline{\varphi}$ and $(\Delta m)_{rms} = -2.5\log \varphi_{rms}$. The linear relation is compatible with the results of \cite{guerras2017}. 

\section{Microlensing Analysis and Constraints on the size of X-Ray Emission Region}

As we can see from Table \ref{obj_tab}, the time spans of the observations $(\Delta t_{obs})$ for our selected targets are sufficiently long, especially when compared to $10R_G$ (typical X-ray source size for radio-quiet quasars) crossing times $(t_{10R_G})$, thus the microlensing light curves span a sufficiently long period to see the typical magnification patterns produced by stars.

Our aim was to obtain probability distributions of the source size for each target individually, by fitting the differential microlensing light curves following \cite{koch2004}. During this process, we used all images for a target. Here, we first generated magnification maps for each image of each target using the three parameters, the dimensionless surface mass density $\kappa$, shear $\gamma$, and fraction of surface density in stars $\kappa_\ast/\kappa$. Since we previously acquired $\kappa$ and $\gamma$ from macrolens models, the last parameter required for generating maps is $\kappa_\ast$. We calculated this parameter from the calibrated relations of \cite{oguri2014} and then we used these values in generating magnification maps with Inverse Polygon Mapping algorithm \citep{media2006}. The lensing parameters are listed in Table \ref{kgtable} including $R/R_{eff}$ (where $R_{eff}$ is the effective radius within which half of the luminosity is emitted), $\kappa_\ast$, $\kappa$ and $\gamma$ values.

\begin{table}[htbp]
  \centering
  \caption{Macrolens Model Parameters of Targets}
    \begin{tabular}{ccccccccc}
    \toprule
    \toprule
    \multirow{2}[4]{*}{Quasar} & \multirow{2}[4]{*}{Image} & \multirow{2}[4]{*}{$R/R_{eff}$} & \multirow{2}[4]{*}{$\kappa_\ast/\kappa$} & \multirow{2}[4]{*}{$\kappa$} & \multirow{2}[4]{*}{$\gamma$} & \multicolumn{3}{c}{Map dimensions} \\
\cmidrule{7-9}          &       &       &       &       &       & Pixels & $R_E$ & $R_G$ \\
    \midrule
    \multirow{4}[2]{*}{MG\,J0414+0534} & A     & 1.617 & 0.288 & 0.489 & 0.454 & \multirow{4}[2]{*}{$4000\times4000$} & \multirow{4}[2]{*}{$19.3\times19.3$} & \multirow{4}[2]{*}{$1500\times1500$} \\
          & B     & 1.582 & 0.296 & 0.530 & 0.524 &       &       &  \\
          & C     & 1.745 & 0.261 & 0.460 & 0.316 &       &       &  \\
          & D     & 1.214 & 0.396 & 0.676 & 0.693 &       &       &  \\
    \midrule
    \multirow{2}[2]{*}{Q\,0957+561} & A     & 2.362 & 0.168 & 0.200 & 0.150 & \multirow{2}[2]{*}{$4000\times4000$} & \multirow{2}[2]{*}{$13.8\times13.8$} & \multirow{2}[2]{*}{$4005\times4005$} \\
          & B     & 0.469 & 0.696 & 1.030 & 0.910 &       &       &  \\
    \midrule
    \multirow{4}[2]{*}{B\,1422+231} & A     & 3.239 & 0.098 & 0.380 & 0.473 & \multirow{4}[2]{*}{$4000\times4000$} & \multirow{4}[2]{*}{$33\times33$} & \multirow{4}[2]{*}{$1500\times1500$} \\
          & B     & 3.095 & 0.106 & 0.492 & 0.628 &       &       &  \\
          & C     & 3.382 & 0.090 & 0.365 & 0.378 &       &       &  \\
          & D     & 0.789 & 0.553 & 1.980 & 2.110 &       &       &  \\
    \bottomrule
    \end{tabular}%
  \label{kgtable}%
\end{table}%

\begin{figure}
	\centering
	\includegraphics[scale=0.45]{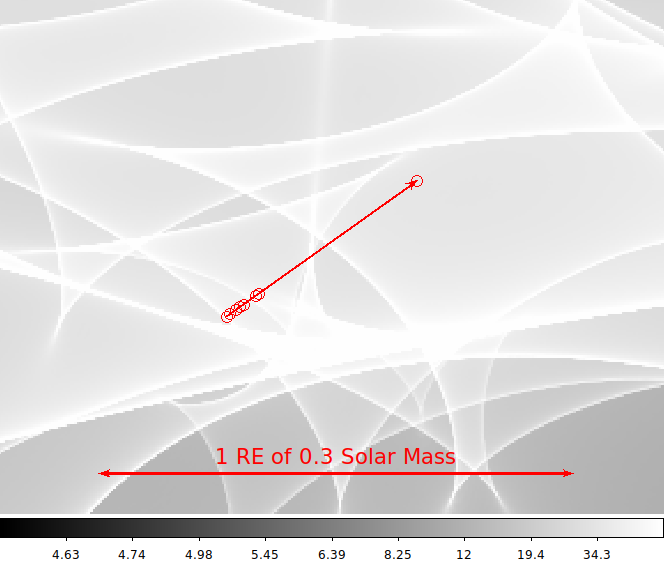}
	\caption{The track which yields the best fit light curves for MG\,J0414+0534 shown on the map of image A. Small circles show the epochs of the actual observations. Darker colours represent smaller magnification.}
	\label{track}
\end{figure}

We took a constant deflector mass of $\langle M_\ast \rangle = 0.3\,M_\odot$ and generated 4000$\times$4000 pixel magnification maps of each image for MG\,J0414+0534 and B\,1422+231, spanning 1500$R_G\times$1500$R_G$ in the source plane. Due to sparsity of caustics for Q\,0957+561, we generated maps with larger pixel sizes but keeping the number of pixels the same, spanning 4005$R_G\times$4005$R_G$ for this target. Considering the values of Einstein radius $(R_E)$ of a $0.3\,M_\odot$ star for each target, the maps span, in the source plane, 155$\times$155 light-days $(19.3\times19.3\,R_E)$ for MG\,J0414+0534, 409$\times$409 light-days $(33\times33\,R_E)$ for B\,1422+231, and 458$\times$458 light-days $(13.8\times13.8\,R_E)$ for Q\,0957+561. We convolved these maps with a Gaussian kernel representing a source model, using the disc surface brightness profile, $I(R) \propto e^{-r^2/R_X^2}$ where $R_X$ is the X-ray source size. Following the work of \cite{guerras2017}, we used a logarithmic grid where $R_X/R_G = e^{0.15n}$ with $n=0,1, 2, \dots, 40$. For each value of $n$, we produced a large number (up to $N=300000$) of simulated light curves choosing randomly oriented tracks on the convolved maps, with lengths equalling the time spans of the observations. An example of these random tracks is shown in Figure \ref{track}. We compared the simulated light curves to the data using $\chi^2$ statistics, where $\chi^2$ for each epoch $t_i$ is
\begin{equation}
	\chi^2 (t_i) = \sum_j \sum_{k<j} \frac{  [\Delta m_{jk}^{obs}(t_i) - \Delta m_{jk}^{sim}(t_i)]^2} {\sigma_{jk}^2(t_i) + \sigma^2(\mu_{jk})}.
\end{equation}
Here $\Delta m_{jk}^{obs}(t_i)$ and $\Delta m_{jk}^{sim}(t_i)$ are the observed and model differential magnitudes respectively at the epoch $t_i$, and $j, k$ represent the images for each lensed quasar. The errors $\sigma_{jk}(t_i) \equiv \sigma_{jk,i}$ are calculated, e.g. for images A and B of a 4-image lensed quasar, using the expression
\begin{equation}
	\frac{1}{\sigma_{AB,i}^2} = \frac{ \sigma_{C,i}^2 \sigma_{D,i}^2 }{ (\sigma_{A,i} \sigma_{B,i}\sigma_{C,i})^2 + (\sigma_{A,i} \sigma_{B,i}\sigma_{D,i})^2 + (\sigma_{A,i} \sigma_{C,i}\sigma_{D,i})^2 + (\sigma_{B,i} \sigma_{C,i}\sigma_{D,i})^2 }
\end{equation}
from \cite{koch2004}, where $\sigma_{j,i}$ are the uncertainties in magnitude units of each image $j$ at each epoch $t_i$. Lastly, $\sigma (\mu_{jk})$ is the uncertainty of the baseline ratio between images $j$ and $k$.

\begin{figure}[!ht]
	\center
	\includegraphics[scale=0.7]{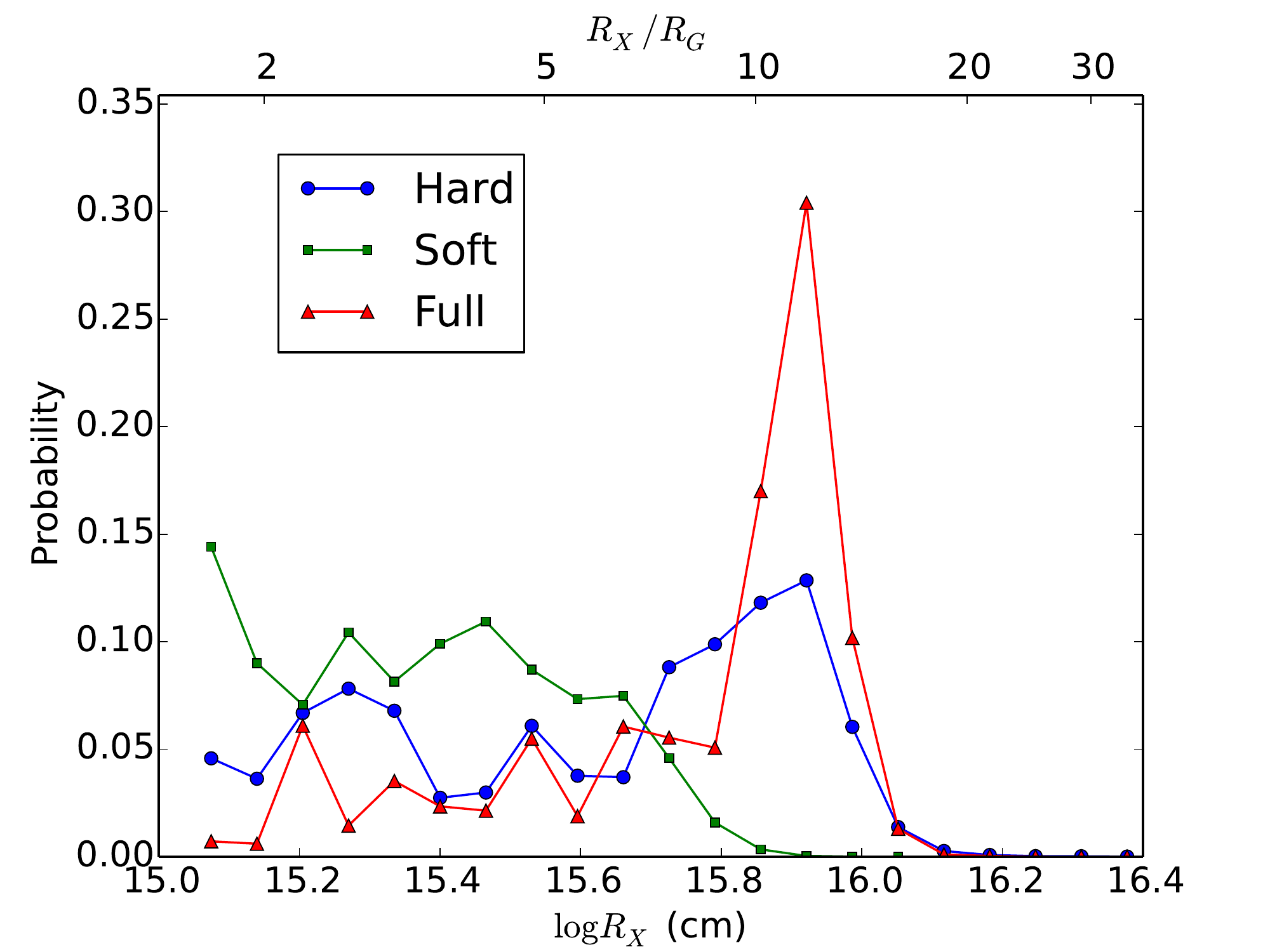}
	\caption{Probability distribution of source size for B\,1422+231}
	\label{bprob}
%
	\center
	\includegraphics[scale=0.7]{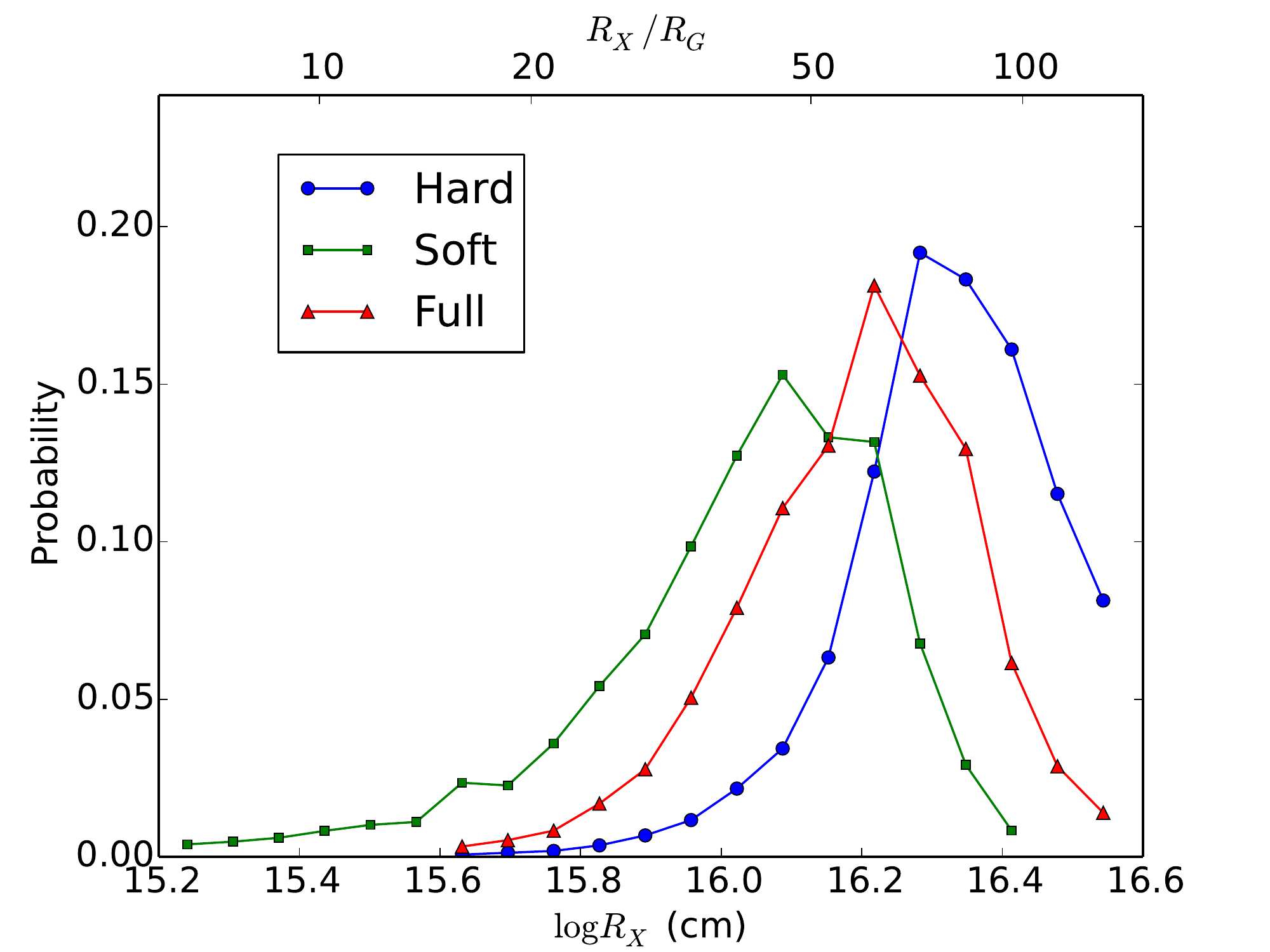}
	\caption{Probability distribution of source size for MG\,J0414+0534}
	\label{mgprob}
\end{figure}

\begin{figure}[!ht]
	\center
	\includegraphics[scale=0.75]{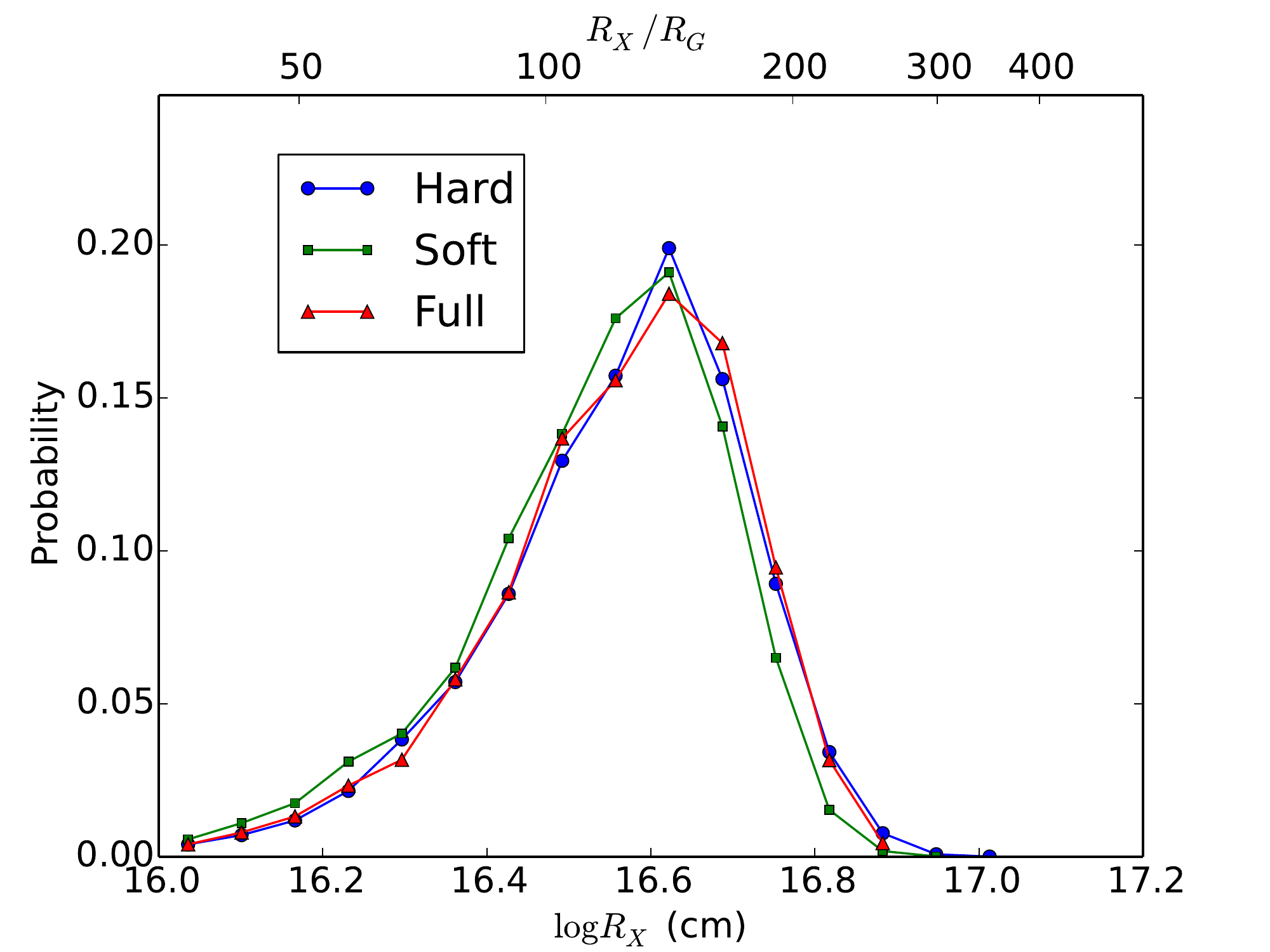}
	\caption{Probability distribution of source size for Q\,0957+561}
	\label{qprob}
\end{figure}

\begin{figure}
	\center
	\includegraphics[scale=0.6]{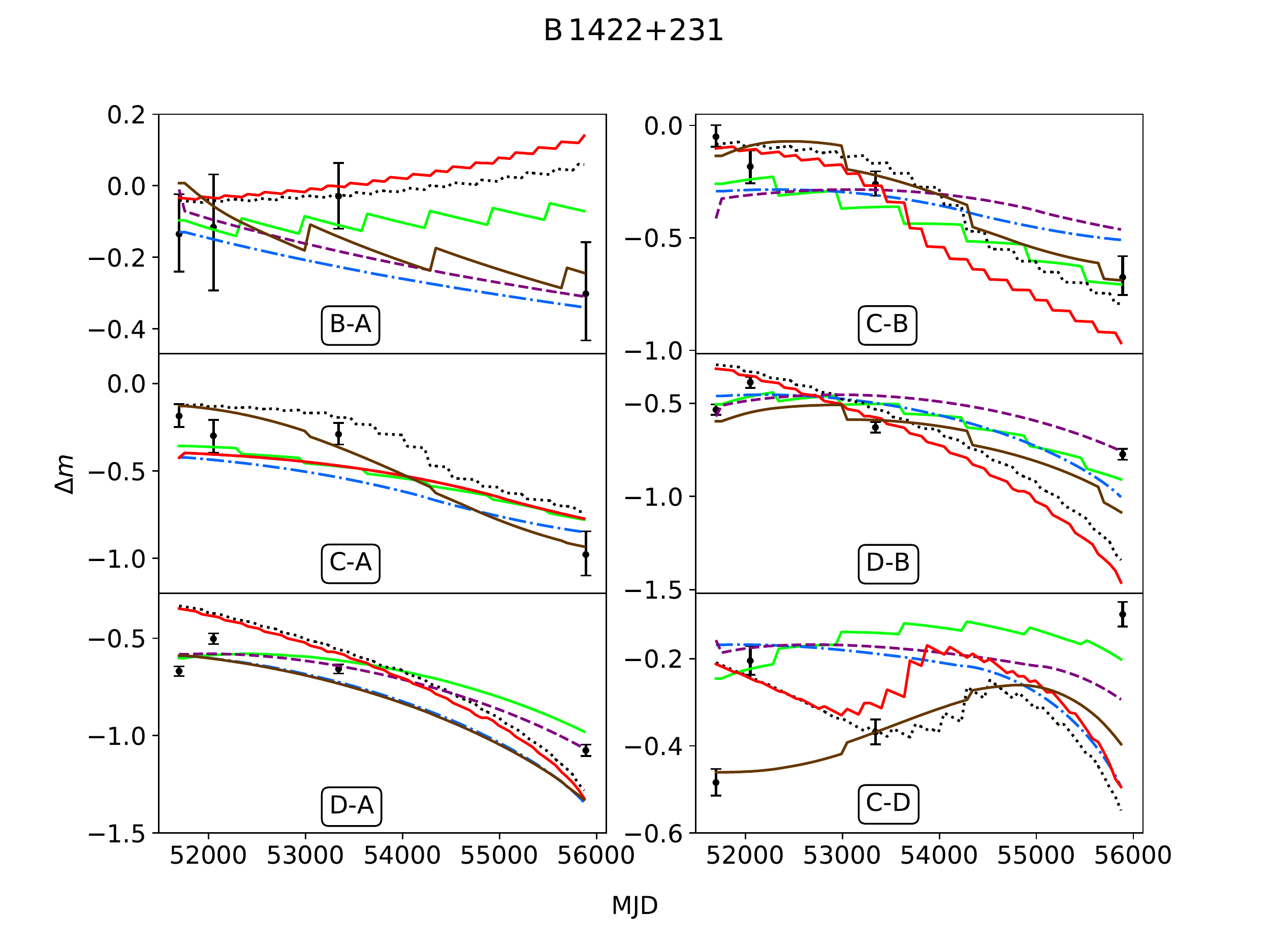}
	\includegraphics[scale=0.6]{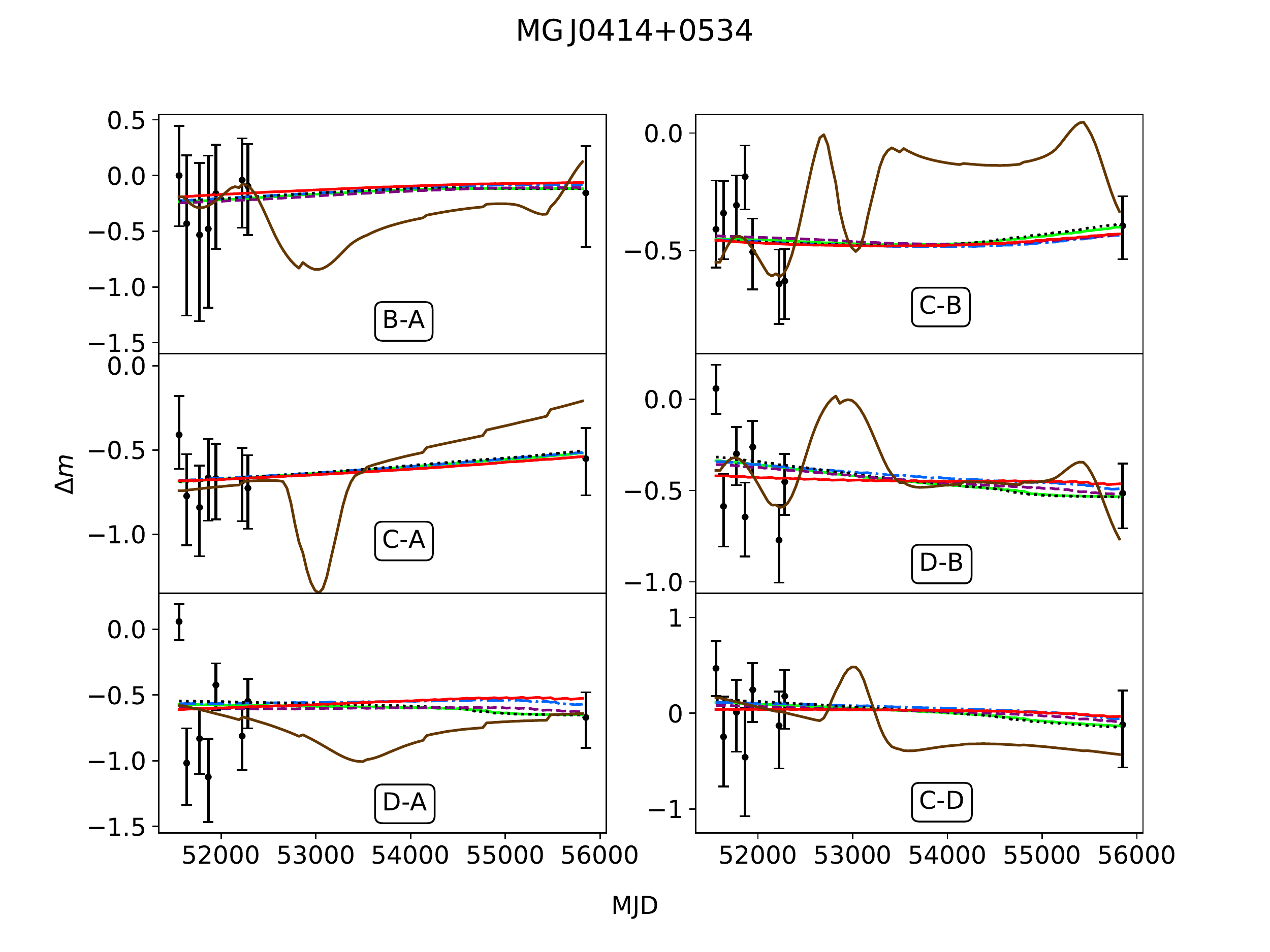}
	\caption{Observed light curves along with the five best fitting models for B\,1422+231 (top) and MG\,J0414+0534 (bottom) taking into account the calculated source sizes. Curves shown in brown represent the minimum $\chi^2$, i.e., source size with maximum likelihood.}
	\label{lcfit1}
\end{figure}

\begin{figure}
	\center
	\includegraphics[scale=0.5]{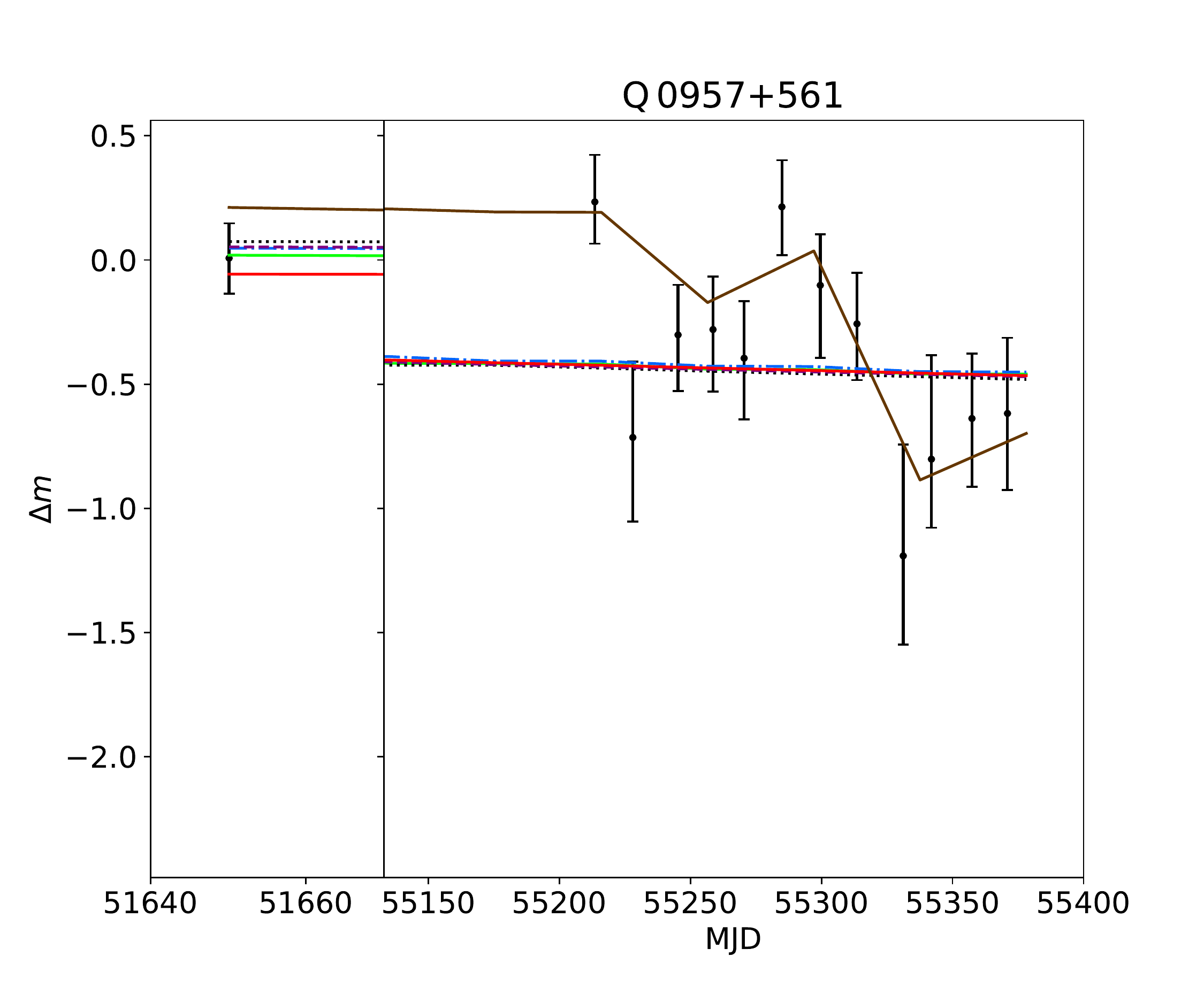}
	\caption{Observed light curves along with the five best fitting models for Q\,0957+561 taking into account the calculated source sizes. The curve shown in brown represents the minimum $\chi^2$, i.e., source size with maximum likelihood.}
	\label{lcfit2}
\end{figure}

For each trial $m$ with a random track on the map, we calculated the likelihood of the source size $R_X$ for each epoch $t_i$ with $L_m(t_i,R_X) = e^{-\chi_m^2(t_i)/2}$, and we acquired the total likelihood for each epoch by adding the likelihoods of all trials,
\begin{equation}
	L(t_i,R_X) = \sum_{m=1}^N e^{-\chi_m^2(t_i)/2}.
\end{equation}
We then obtained the probability of the differential microlensing amplitude $\Delta m_{jk}$ for a particular source size $R_X$ by multiplying the likelihoods of all epochs, 
\begin{equation}
	p(\Delta m_{jk} | R_X) = \prod_{t_i} L(t_i,R_X)
	\label{probrx}
\end{equation}
After obtaining the probabilities $p(\Delta m_{jk} | R_X)$ for each source size, we normalised them by their sum and plotted against the source size. Finally, we acquired the size estimates by fitting each probability distribution with a Gaussian. Probability distributions are shown in Figure \ref{bprob}--\ref{qprob} whereas the size estimates, assuming a ``face-on disc" in which the inclination angle of the disc is $i=0\degree$, are given in table \ref{qso_size}. If the disc is not viewed face-on, these estimates will scale as $(\cos i)^{-1/2}$ \citep{dai2010}. Finally, in Figures \ref{lcfit1} and \ref{lcfit2}, we present a sample of best-fit light curves taking into account the obtained $R_X$ values.

\begin{table}[htbp]
  \centering
  \caption{X-ray Source Size Estimates With Bayesian Probabilities}
    \begin{tabular}{lcccccc}
    \toprule
    \toprule
    \multicolumn{1}{c}{Quasar} & $\log (R_X^{soft}/cm)$ & $\log (R_X^{hard}/cm)$ & $\log (R_X^{full}/cm)$ & $R_X^{soft}/R_G$ & $R_X^{hard}/R_G$ & $R_X^{full}/R_G$ \\
    \midrule
    MG\,J0414+0534 & $16.08\pm0.17$ & $16.34\pm0.14$ & $16.22\pm0.16$ & $45.30\pm17.75$ & $82.28\pm26.73$ & $61.27\pm21.87$ \\
    Q\,0957+561 & $16.57\pm0.14$ & $16.59\pm0.14$ & $16.59\pm0.14$ & $125.54\pm39.15$ & $132.02\pm41.95$ & $132.11\pm42.82$ \\
    B\,1422+231 & $15.22\pm0.37$ & $15.64\pm0.39$ & $15.91\pm0.05$ & $2.34\pm1.97$ & $6.17\pm5.48$ & $11.51\pm1.42$ \\
    \bottomrule
    \end{tabular}
    \label{qso_size}
\end{table}

\begin{figure}[!ht]
	\center
	\includegraphics[scale=0.75]{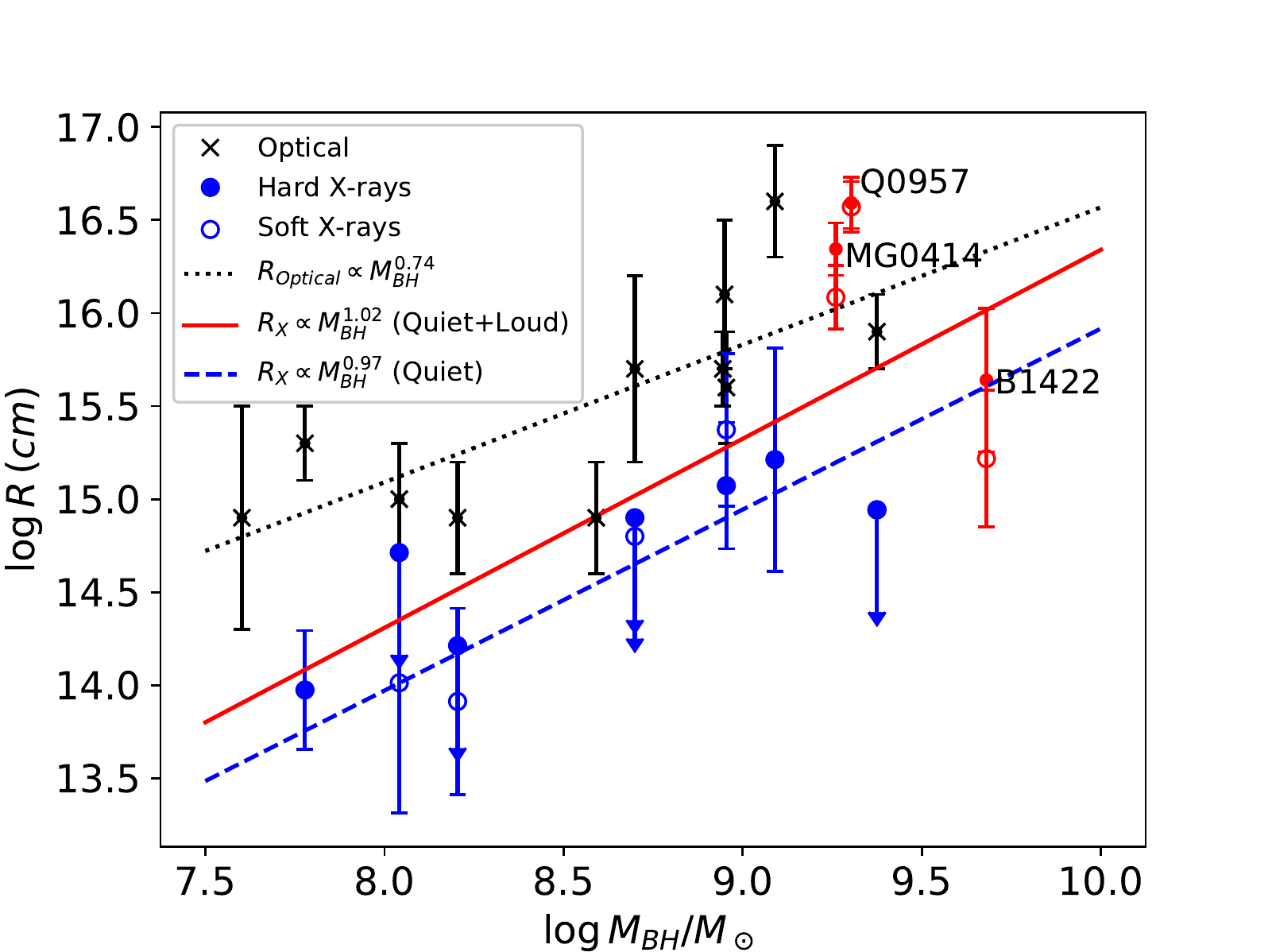}
	\caption{Accretion disc sizes plotted against black hole mass}
	\label{sizecomp}
\end{figure}

\begin{figure}
	\centering
	\includegraphics[scale=0.6]{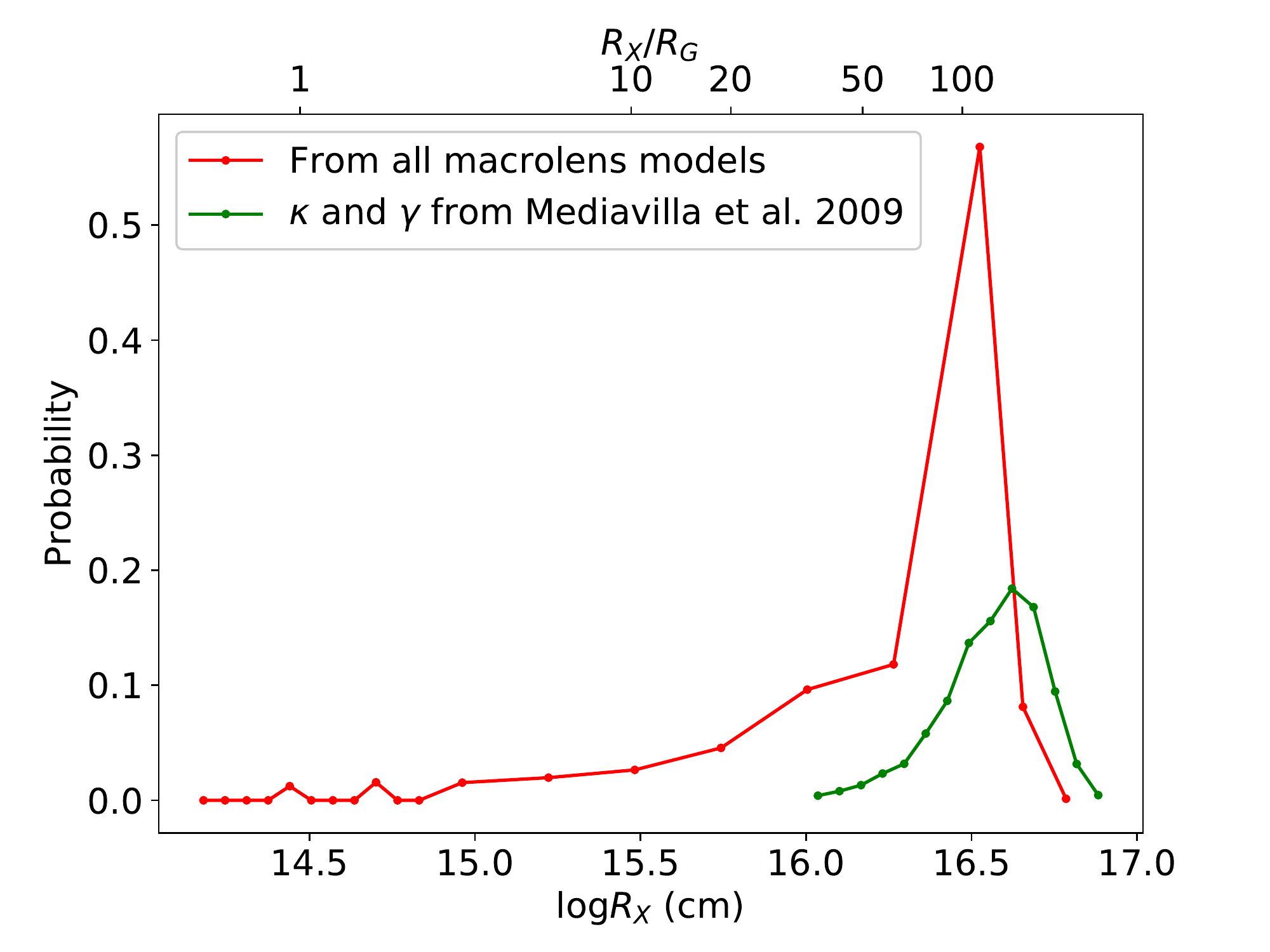}
	\caption{Probability distributions of source size in full band for Q\,0957+561.}
	\label{prob_comp}
\end{figure}

We also calculated the source size for Q\,0957+561 in full band considering different macro models, which are described by the fraction of mass in the de Vaucouleurs component $(f_\ast)$. We took models with $0.1\le f_\ast \le 1$ in equal steps, generated maps with $\kappa$ and $\gamma$ corresponding to these $f_\ast$ values, and calculated the probability distribution of source size from simulated light curves. Here, we obtained the probability for a particular source size $R_X$ by summing the probabilities from all $f_\ast$ values. Accordingly, source size was calculated to be $\log R_X^{full}/cm=16.45 \pm 0.10$, which is in accordance with the value $\log R_X^{full}/cm=16.59 \pm 0.14$ given in Table \ref{qso_size}. The probability distribution obtained by considering all macro models, and the one obtained by taking the macro parameters from \cite{media2009} are given in Figure \ref{prob_comp}. Finally, for the hard band, we calculated the source sizes which have the maximum likelihood, i.e., which correspond to the best fit light curves with the lowest $\chi^2$. We give the resulting source sizes in Table \ref{maxlike}, and best fit light curves corresponding to those sizes in Figures \ref{lcfit1} and \ref{lcfit2}.

\begin{table}[htbp]
  \centering
  \caption{Source Sizes With Maximum Likelihood}
    \begin{tabular}{lccc}
    \toprule
    \toprule
    \multicolumn{1}{c}{Quasar} & $R_X^{soft}/R_G$ & $R_X^{hard}/R_G$ & $R_X^{full}/R_G$ \\
    \midrule
    MG\,J0414+0534 & $21.52^{+26.82}_{-20.52}$ & $1.45^{+1.79}_{-0.45}$ & $15.95^{+38.03}_{-14.95}$ \\
    Q\,0957+561 & $2.46\pm0.60$ & $2.46\pm0.60$ & $2.12\pm0.47$ \\
    B\,1422+231 & $10.17\pm5.05$ & $11.81\pm3.75$ & $21.52\pm2.65$ \\
    \bottomrule
    \end{tabular}
  \label{maxlike}
\end{table}

\begin{figure}
	\centering
	\includegraphics{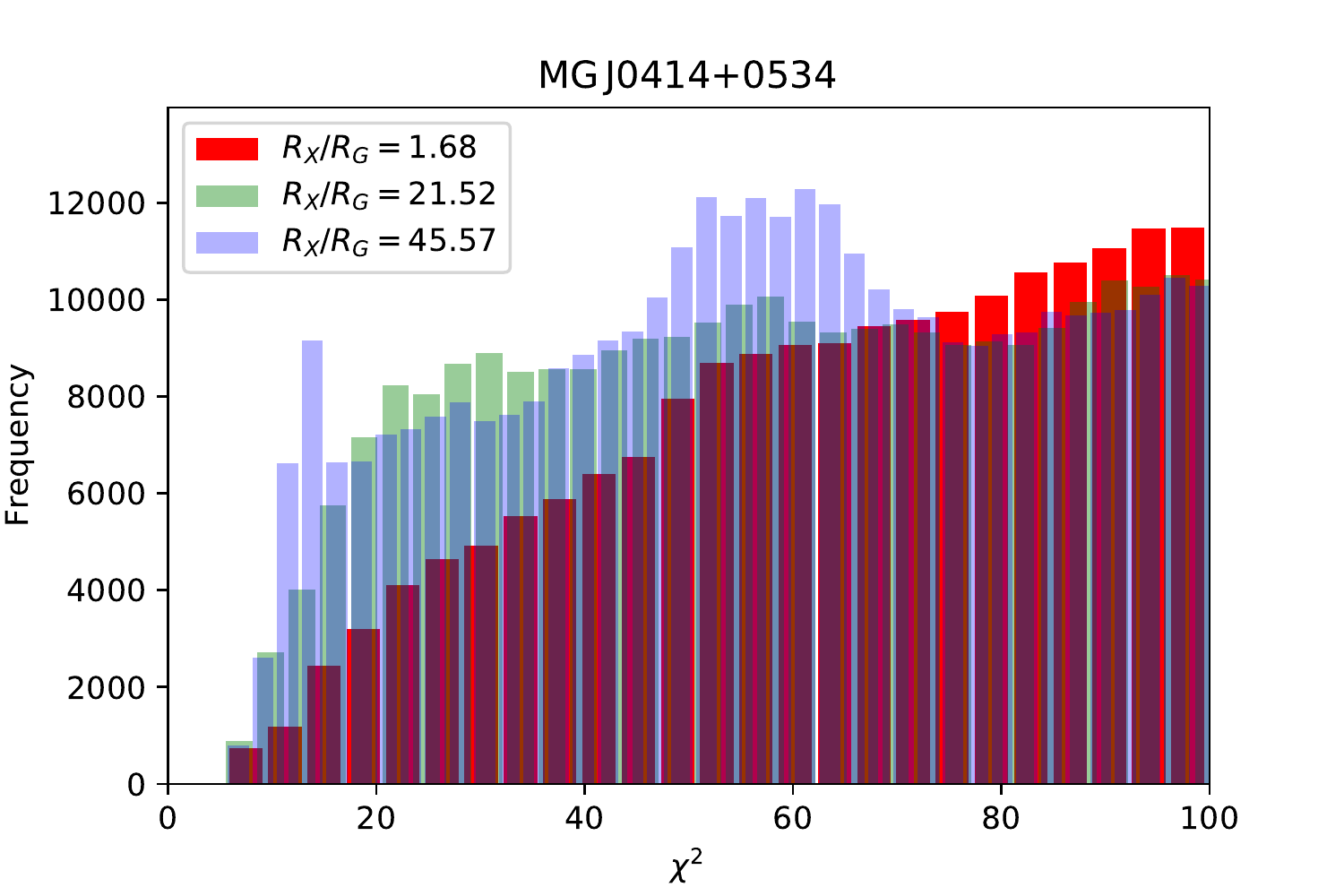}
	\caption{$\chi^2$ distribution for $10^6$ trials on maps of MG\,J0414+0534 for $\chi^2<100$}
	\label{mgchi}
\end{figure}
\begin{figure}
	\centering
	\includegraphics{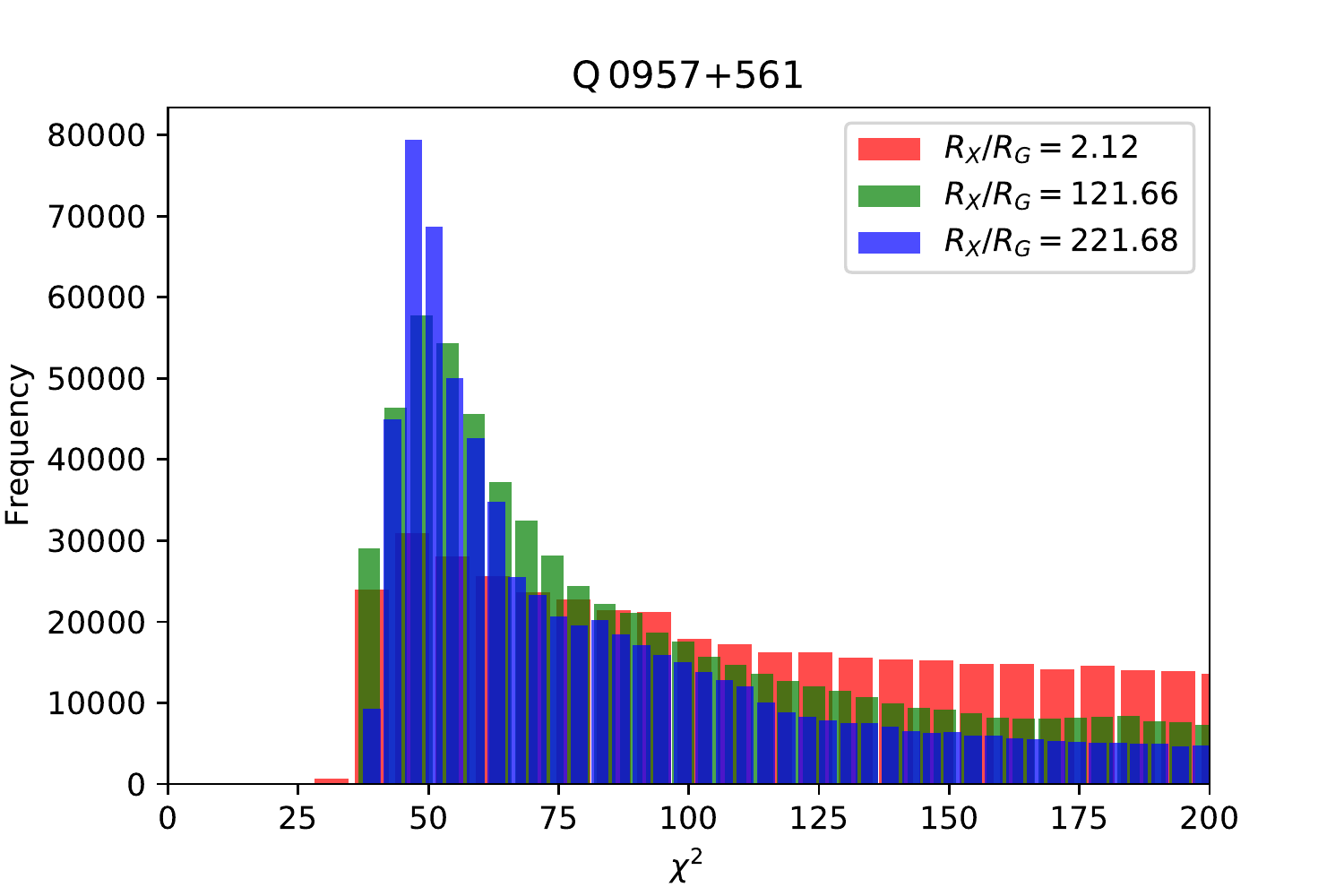}
	\caption{$\chi^2$ distribution for $10^6$ trials on maps of Q\,0957+561 for $\chi^2<200$}
	\label{qchi}
\end{figure}
\begin{figure}
	\centering
	\includegraphics{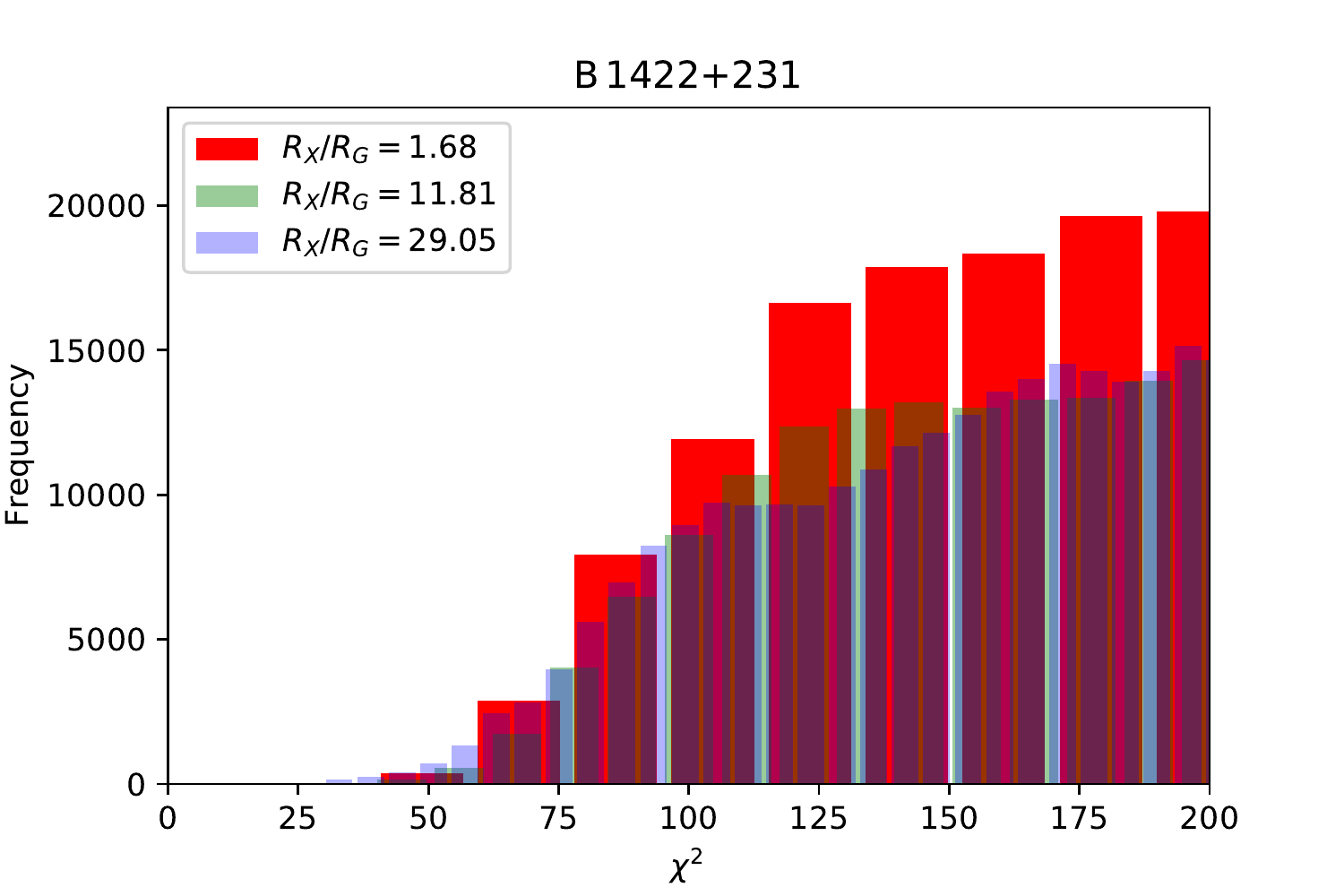}
	\caption{$\chi^2$ distribution for $10^6$ trials on maps of B\,1422+231 for $\chi^2<200$}
	\label{bchi}
\end{figure}

To compare our results with the sizes of other lensed quasars in UV and X-ray bands, we used the data given in \cite{morgan2010} and plotted the accretion disc sizes against black hole mass (Figure \ref{sizecomp}). As seen in Figure \ref{sizecomp}, our size estimates are in agreement with the apparent relation between the X-ray source size and the black hole mass, roughly as $R_X \propto M_{BH}$. These results also imply that the radio-loud quasars tend to have larger X-ray emission regions compared to radio-quiet quasars. In an effort to understand the origin of this difference, we also examine the rms of microlensing variability. From Figure \ref{rms}, we could see that radio-quiet quasars HE\,0435-1223 and QJ\,0158-4325 have very similar microlensing amplitudes to the ones in our sample. However, their hard X-ray region sizes are $\log R_X/cm=14.9$ \citep{blackburne2014} and 14.21 \citep{morgan2012} respectively, which are significantly smaller than the ones for our radio-loud sample. The fact that the Bayesian and maximum likelihood sizes are not consistent with each other, except for B\,1422+231, suggests that more data with better signal-to-noise ratio are needed to better constrain the sizes of MG\,J0414+0534 and Q\,0957+561. Furthermore, X-ray region sizes of MG\,J0414+0534, Q\,0957+561 are greater than their Einstein radii, whereas, in case of the two radio-quiet quasars with similar microlensing amplitudes as mentioned above, X-ray sizes are much smaller than their Einstein radii. This fact that the resulting source sizes are very different despite the similar microlensing amplitudes, raises even more questions. As seen in Figures \ref{lcfit1} and \ref{lcfit2}, model light curves do not fit the small fluctuations, which  possibly provides an explanation for why the data yield large values of rms of microlensing variability despite the large source size. Besides, as expected for large source size, when rms is calculated from the model, they are much smaller than the ones calculated from the observations. Lastly, as seen from Figures \ref{mgchi} -- \ref{bchi}, even though there are light curve solutions from small source sizes with lower $\chi^2$ values, these are very few in numbers. However, large source sizes dominantly contribute to probability with slightly bigger $\chi^2$ values, which explains the large source sizes being much more probable even though their light curves do not fit the small fluctuations well. Obviously, the fact that smaller $\chi^2$ values are achieved with smaller $R_X$ also explains why the maximum likelihood source sizes are extremely small (apart from B\,1422+231) compared to the results from Bayesian analysis.

\section{Discussion and Conclusion} \label{conc}

In this paper, we present the X-ray monitoring results of three lensed radio-loud quasars MG\,J0414+0534, Q\,0957+561 and B\,1422+231. We performed both spectroscopic and photometric analysis of \textit{Chandra} archival data. In our spectroscopic analysis, we found that a power law model modified by absorption with additional Gaussian emission lines provide good fits to spectral data. As a result of these fits, we tentatively detected the characteristic FeK$\alpha$ line in MG\,J0414+0534 and Q\,0957+561 with over 95\% significance. 

FeK$\alpha$ line shifts detected in our spectral analysis might be caused by a caustic passing through the inner accretion disc as discussed by \cite{chartas2012}. The two lines in image B of Q\,0957+561 can be new examples of the distortions of a single FeK$\alpha$ line due to special relativistic Doppler and general relativistic effects, then magnified by microlensing. For radio-quiet quasars, as concluded by \cite{chartas2017}, these shifts in FeK$\alpha$ line energy is formed by reflection from the material near the black hole horizon because of the small X-ray corona size. Here, our Bayesian microlensing X-ray size for Q\,0957+561 is much larger. Assuming little general relativistic effects, Doppler shifted FeK$\alpha$ line energy calculated with the source size given in Table \ref{qso_size} can reach $7.31\pm0.26$ keV when magnified by a microlensing caustic, which is in fact compatible with the observed line energies in both images.

We also obtained microlensing light curves from flux ratios measured from PSF fitting of the absorption corrected data. As seen in Figures \ref{lc0414}--\ref{lc1422}, there is no significant difference in flux ratios between soft and hard X-ray bands, i.e. an energy dependent microlensing, apart from the C image of MG\,J0414+0534 at modified Julian date around 52000, which needs to be further confirmed with more observations.

From the size estimates given in table \ref{qso_size}, we calculated the size ratios of soft and hard as $\log (R_X^{hard}/R_X^{soft}) = 0.26 \pm 0.22,\, 0.02 \pm 0.19$, $0.42 \pm 0.53$ for MG\,J0414+0534, Q\,0957+561 and B\,1422+231 respectively. These values do not support the intuitive idea of the hard component being more compact than the soft one, towards which also \cite{mosq2013} could not find a strong evidence.

Our X-ray microlensing analysis results for Q\,0957+561 suggest a much smaller X-ray source size compared to the X-ray/UV and optical reverberation mapping results from \cite{gilmer2012} in which they found $R_X \sim 200\,R_S\sim 0.05$ pc (with $M_{BH}=2.5\times 10^9\,M_\odot$), whereas our result is $R_X\sim 65\,R_S\sim 0.0125$ pc (average of soft, hard and full). To put stricter constraints on X-ray source sizes, we need more data with higher signal-to-noise ratio which will make it possible to have light curves with smaller error bars.

\acknowledgments

%

\vspace{5mm}
\facilities{\textit{Chandra} X-Ray Satellite}


\software{XSPEC \citep{arnaud1996}}

\clearpage
\bibliography{ref}

\begin{thebibliography}{}
\expandafter\ifx\csname natexlab\endcsname\relax\def\natexlab#1{#1}\fi
\providecommand{\url}[1]{\href{#1}{#1}}
\providecommand{\dodoi}[1]{doi:~\href{http://doi.org/#1}{\nolinkurl{#1}}}
\providecommand{\doeprint}[1]{\href{http://ascl.net/#1}{\nolinkurl{http://ascl.net/#1}}}
\providecommand{\doarXiv}[1]{\href{https://arxiv.org/abs/#1}{\nolinkurl{https://arxiv.org/abs/#1}}}

\bibitem[{{Arnaud}(1996)}]{arnaud1996}
{Arnaud}, K.~A. 1996, in Astronomical Society of the Pacific Conference Series,
  Vol. 101, Astronomical Data Analysis Software and Systems V, ed. G.~H.
  {Jacoby} \& J.~{Barnes}, 17

\bibitem[{{Blackburne} {et~al.}(2014){Blackburne}, {Kochanek}, {Chen}, {Dai},
  \& {Chartas}}]{blackburne2014}
{Blackburne}, J.~A., {Kochanek}, C.~S., {Chen}, B., {Dai}, X., \& {Chartas}, G.
  2014, \apj, 789, 125, \dodoi{10.1088/0004-637X/789/2/125}

\bibitem[{{Blaes}(2007)}]{blaes2007}
{Blaes}, O. 2007, in Astronomical Society of the Pacific Conference Series,
  Vol. 373, The Central Engine of Active Galactic Nuclei, ed. L.~C. {Ho} \&
  J.-W. {Wang}, 75

\bibitem[{{Chartas} {et~al.}(2002){Chartas}, {Agol}, {Eracleous}, {Garmire},
  {Bautz}, \& {Morgan}}]{chartas2002}
{Chartas}, G., {Agol}, E., {Eracleous}, M., {et~al.} 2002, \apj, 568, 509,
  \dodoi{10.1086/339162}

\bibitem[{{Chartas} {et~al.}(2012){Chartas}, {Kochanek}, {Dai}, {Moore},
  {Mosquera}, \& {Blackburne}}]{chartas2012}
{Chartas}, G., {Kochanek}, C.~S., {Dai}, X., {et~al.} 2012, \apj, 757, 137,
  \dodoi{10.1088/0004-637X/757/2/137}

\bibitem[{{Chartas} {et~al.}(2017){Chartas}, {Krawczynski}, {Zalesky},
  {Kochanek}, {Dai}, {Morgan}, \& {Mosquera}}]{chartas2017}
{Chartas}, G., {Krawczynski}, H., {Zalesky}, L., {et~al.} 2017, \apj, 837, 26,
  \dodoi{10.3847/1538-4357/aa5d50}

\bibitem[{{Chartas} {et~al.}(2000){Chartas}, {Worrall}, {Birkinshaw},
  {Cresitello-Dittmar}, {Cui}, {Ghosh}, {Harris}, {Hooper}, {Jauncey}, {Kim},
  {Lovell}, {Mathur}, {Schwartz}, {Tingay}, {Virani}, \&
  {Wilkes}}]{chartas2000}
{Chartas}, G., {Worrall}, D.~M., {Birkinshaw}, M., {et~al.} 2000, \apj, 542,
  655, \dodoi{10.1086/317049}

\bibitem[{{Chen} {et~al.}(2012){Chen}, {Dai}, {Kochanek}, {Chartas},
  {Blackburne}, \& {Morgan}}]{chen2012}
{Chen}, B., {Dai}, X., {Kochanek}, C.~S., {et~al.} 2012, \apj, 755, 24,
  \dodoi{10.1088/0004-637X/755/1/24}

\bibitem[{{Dai} {et~al.}(2003){Dai}, {Chartas}, {Agol}, {Bautz}, \&
  {Garmire}}]{dai2003}
{Dai}, X., {Chartas}, G., {Agol}, E., {Bautz}, M.~W., \& {Garmire}, G.~P. 2003,
  \apj, 589, 100, \dodoi{10.1086/374548}

\bibitem[{{Dai} \& {Guerras}(2018)}]{dai2018}
{Dai}, X., \& {Guerras}, E. 2018, ApJ Letters, 853, L27,
  \dodoi{10.3847/2041-8213/aaa5fb}

\bibitem[{{Dai} {et~al.}(2010){Dai}, {Kochanek}, {Chartas}, {Koz{\l}owski},
  {Morgan}, {Garmire}, \& {Agol}}]{dai2010}
{Dai}, X., {Kochanek}, C.~S., {Chartas}, G., {et~al.} 2010, \apj, 709, 278,
  \dodoi{10.1088/0004-637X/709/1/278}

\bibitem[{{Dai} {et~al.}(2019){Dai}, {Steele}, {Guerras}, {Morgan}, \&
  {Chen}}]{dai2019}
{Dai}, X., {Steele}, S., {Guerras}, E., {Morgan}, C.~W., \& {Chen}, B. 2019,
  arXiv e-prints, arXiv:1901.06007.
\newblock \doarXiv{1901.06007}

\bibitem[{{Dickey} \& {Lockman}(1990)}]{dickey1990}
{Dickey}, J.~M., \& {Lockman}, F.~J. 1990, \araa, 28, 215,
  \dodoi{10.1146/annurev.aa.28.090190.001243}

\bibitem[{{Fabian} {et~al.}(1995){Fabian}, {Nandra}, {Reynolds}, {Brandt},
  {Otani}, {Tanaka}, {Inoue}, \& {Iwasawa}}]{fabian1995}
{Fabian}, A.~C., {Nandra}, K., {Reynolds}, C.~S., {et~al.} 1995, \mnras, 277,
  L11, \dodoi{10.1093/mnras/277.1.L11}

\bibitem[{{Ferrarese} \& {Merritt}(2000)}]{fer2000}
{Ferrarese}, L., \& {Merritt}, D. 2000, \apjl, 539, L9, \dodoi{10.1086/312838}

\bibitem[{{George} \& {Fabian}(1991)}]{georgefabian1991}
{George}, I.~M., \& {Fabian}, A.~C. 1991, \mnras, 249, 352,
  \dodoi{10.1093/mnras/249.2.352}

\bibitem[{{Gil-Merino} {et~al.}(2012){Gil-Merino}, {Goicoechea}, {Shalyapin},
  \& {Braga}}]{gilmer2012}
{Gil-Merino}, R., {Goicoechea}, L.~J., {Shalyapin}, V.~N., \& {Braga}, V.~F.
  2012, \apj, 744, 47, \dodoi{10.1088/0004-637X/744/1/47}

\bibitem[{{Gou} {et~al.}(2011){Gou}, {McClintock}, {Reid}, {Orosz}, {Steiner},
  {Narayan}, {Xiang}, {Remillard}, {Arnaud}, \& {Davis}}]{gou2011}
{Gou}, L., {McClintock}, J.~E., {Reid}, M.~J., {et~al.} 2011, \apj, 742, 85,
  \dodoi{10.1088/0004-637X/742/2/85}

\bibitem[{{Guerras} {et~al.}(2018){Guerras}, {Dai}, \&
  {Mediavilla}}]{guerras2018}
{Guerras}, E., {Dai}, X., \& {Mediavilla}, E. 2018, arXiv e-prints,
  arXiv:1805.11498.
\newblock \doarXiv{1805.11498}

\bibitem[{{Guerras} {et~al.}(2017){Guerras}, {Dai}, {Steele}, {Liu},
  {Kochanek}, {Chartas}, {Morgan}, \& {Chen}}]{guerras2017}
{Guerras}, E., {Dai}, X., {Steele}, S., {et~al.} 2017, \apj, 836, 206,
  \dodoi{10.3847/1538-4357/aa5728}

\bibitem[{{Jiang} {et~al.}(2007){Jiang}, {Fan}, {Ivezi{\'c}}, {Richards},
  {Schneider}, {Strauss}, \& {Kelly}}]{jiang2007}
{Jiang}, L., {Fan}, X., {Ivezi{\'c}}, {\v Z}., {et~al.} 2007, \apj, 656, 680,
  \dodoi{10.1086/510831}

\bibitem[{{Kellermann} {et~al.}(1989){Kellermann}, {Sramek}, {Schmidt},
  {Shaffer}, \& {Green}}]{keller1989}
{Kellermann}, K.~I., {Sramek}, R., {Schmidt}, M., {Shaffer}, D.~B., \& {Green},
  R. 1989, \aj, 98, 1195, \dodoi{10.1086/115207}

\bibitem[{{Kochanek}(2004)}]{koch2004}
{Kochanek}, C.~S. 2004, ApJ, 605, 58, \dodoi{10.1086/382180}

\bibitem[{{Kormendy} \& {Richstone}(1995)}]{kormendy1995}
{Kormendy}, J., \& {Richstone}, D. 1995, \araa, 33, 581,
  \dodoi{10.1146/annurev.aa.33.090195.003053}

\bibitem[{{Marshall} {et~al.}(2018){Marshall}, {Gelbord}, {Worrall},
  {Birkinshaw}, {Schwartz}, {Jauncey}, {Griffiths}, {Murphy}, {Lovell},
  {Perlman}, \& {Godfrey}}]{marshall2018}
{Marshall}, H.~L., {Gelbord}, J.~M., {Worrall}, D.~M., {et~al.} 2018, \apj,
  856, 66, \dodoi{10.3847/1538-4357/aaaf66}

\bibitem[{{McConnell} \& {Ma}(2013)}]{mccon2013}
{McConnell}, N.~J., \& {Ma}, C.-P. 2013, \apj, 764, 184,
  \dodoi{10.1088/0004-637X/764/2/184}

\bibitem[{{Mediavilla} {et~al.}(2006){Mediavilla}, {Mu{\~n}oz}, {Lopez},
  {Mediavilla}, {Abajas}, {Gonzalez-Morcillo}, \& {Gil-Merino}}]{media2006}
{Mediavilla}, E., {Mu{\~n}oz}, J.~A., {Lopez}, P., {et~al.} 2006, \apj, 653,
  942, \dodoi{10.1086/508796}

\bibitem[{{Mediavilla} {et~al.}(2009){Mediavilla}, {Mu{\~n}oz}, {Falco},
  {Motta}, {Guerras}, {Canovas}, {Jean}, {Oscoz}, \& {Mosquera}}]{media2009}
{Mediavilla}, E., {Mu{\~n}oz}, J.~A., {Falco}, E., {et~al.} 2009, \apj, 706,
  1451, \dodoi{10.1088/0004-637X/706/2/1451}

\bibitem[{{Morgan} {et~al.}(2010){Morgan}, {Kochanek}, {Morgan}, \&
  {Falco}}]{morgan2010}
{Morgan}, C.~W., {Kochanek}, C.~S., {Morgan}, N.~D., \& {Falco}, E.~E. 2010,
  \apj, 712, 1129, \dodoi{10.1088/0004-637X/712/2/1129}

\bibitem[{{Morgan} {et~al.}(2012){Morgan}, {Hainline}, {Chen}, {Tewes},
  {Kochanek}, {Dai}, {Kozlowski}, {Blackburne}, {Mosquera}, {Chartas},
  {Courbin}, \& {Meylan}}]{morgan2012}
{Morgan}, C.~W., {Hainline}, L.~J., {Chen}, B., {et~al.} 2012, \apj, 756, 52,
  \dodoi{10.1088/0004-637X/756/1/52}

\bibitem[{{Mosquera} \& {Kochanek}(2011)}]{mosqkoch2011}
{Mosquera}, A.~M., \& {Kochanek}, C.~S. 2011, \apj, 738, 96,
  \dodoi{10.1088/0004-637X/738/1/96}

\bibitem[{{Mosquera} {et~al.}(2013){Mosquera}, {Kochanek}, {Chen}, {Dai},
  {Blackburne}, \& {Chartas}}]{mosq2013}
{Mosquera}, A.~M., {Kochanek}, C.~S., {Chen}, B., {et~al.} 2013, ApJ, 769, 53,
  \dodoi{10.1088/0004-637X/769/1/53}

\bibitem[{{Oguri} {et~al.}(2014){Oguri}, {Rusu}, \& {Falco}}]{oguri2014}
{Oguri}, M., {Rusu}, C.~E., \& {Falco}, E.~E. 2014, \mnras, 439, 2494,
  \dodoi{10.1093/mnras/stu106}

\bibitem[{{Protassov} {et~al.}(2002){Protassov}, {van Dyk}, {Connors},
  {Kashyap}, \& {Siemiginowska}}]{protas2002}
{Protassov}, R., {van Dyk}, D.~A., {Connors}, A., {Kashyap}, V.~L., \&
  {Siemiginowska}, A. 2002, \apj, 571, 545, \dodoi{10.1086/339856}

\bibitem[{{Schechter} {et~al.}(2014){Schechter}, {Pooley}, {Blackburne}, \&
  {Wambsganss}}]{sch2014}
{Schechter}, P.~L., {Pooley}, D., {Blackburne}, J.~A., \& {Wambsganss}, J.
  2014, \apj, 793, 96, \dodoi{10.1088/0004-637X/793/2/96}

\bibitem[{{Schwartz} {et~al.}(2000){Schwartz}, {Marshall}, {Lovell}, {Piner},
  {Tingay}, {Birkinshaw}, {Chartas}, {Elvis}, {Feigelson}, {Ghosh}, {Harris},
  {Hirabayashi}, {Hooper}, {Jauncey}, {Lanzetta}, {Mathur}, {Preston},
  {Tucker}, {Virani}, {Wilkes}, \& {Worrall}}]{schwartz2000}
{Schwartz}, D.~A., {Marshall}, H.~L., {Lovell}, J.~E.~J., {et~al.} 2000, \apjl,
  540, 69, \dodoi{10.1086/312875}

\bibitem[{{Somerville} {et~al.}(2008){Somerville}, {Hopkins}, {Cox},
  {Robertson}, \& {Hernquist}}]{somer2008}
{Somerville}, R.~S., {Hopkins}, P.~F., {Cox}, T.~J., {Robertson}, B.~E., \&
  {Hernquist}, L. 2008, \mnras, 391, 481,
  \dodoi{10.1111/j.1365-2966.2008.13805.x}

\bibitem[{{Urry} \& {Padovani}(1995)}]{urry1995}
{Urry}, C.~M., \& {Padovani}, P. 1995, \pasp, 107, 803, \dodoi{10.1086/133630}

\bibitem[{{Walsh} {et~al.}(1979){Walsh}, {Carswell}, \& {Weymann}}]{walsh1979}
{Walsh}, D., {Carswell}, R.~F., \& {Weymann}, R.~J. 1979, Nature, 279, 381,
  \dodoi{10.1038/279381a0}

\bibitem[{{Wambsganss}(2006)}]{wamb2006}
{Wambsganss}, J. 2006, in Saas-Fee Advanced Course 33: Gravitational Lensing:
  Strong, Weak and Micro, ed. G.~{Meylan}, P.~{Jetzer}, P.~{North},
  P.~{Schneider}, C.~S. {Kochanek}, \& J.~{Wambsganss}, 453--540

\bibitem[{{Wilson} \& {Colbert}(1995)}]{wilson1995}
{Wilson}, A.~S., \& {Colbert}, E.~J.~M. 1995, \apj, 438, 62,
  \dodoi{10.1086/175054}

\end{thebibliography}
\end{document}